\documentclass[%
 aip,
 amsmath,amssymb,
 preprint,%
]{revtex4-1}

\usepackage{graphicx}
\usepackage{dcolumn}
\usepackage{bm}

\usepackage[utf8]{inputenc}
\usepackage[T1]{fontenc}
\usepackage{mathptmx}

\begin{document}



\author{Moritz Heindl}
\author{Leticia Gonz\'alez}
 \email{leticia.gonzalez@univie.ac.at}
\affiliation{Institute of Theoretical Chemistry, Faculty of Chemistry, University of Vienna, Währingerstr. 17, 1090 Vienna, Austria
}

\date{\today}

\noindent
{ The following article has been submitted to the Journal of Chemical Physics. After it is published, it will be found at https://aip.scitation.org/journal/jcp} 
\vspace{2cm}

\begin{center}
    {\bf \Large Validating Fewest-Switches Surface Hopping in the Presence of Laser Fields}

\end{center}

\begin{abstract}
The capability of fewest-switches surface hopping (FSSH) to describe non-adiabatic dynamics of small and medium sized molecules under explicit excitation with external fields 
is evaluated.
Different parameters in FSSH and combinations thereof are benchmarked against multi-configurational time dependent Hartree (MCTDH) reference calculations using SO$_2$ and 2-thiocytosine as model, yet realistic, molecular systems. 
Qualitatively, we find that FSSH is able to reproduce the trends in the MCTDH dynamics with (and without) an explicit external field; however, no set of FSSH parameters is ideal.
An adequate treatment of the overcoherence in FSSH is identified as the driving factor to improve the description of the excitation process with respect to the MCTDH reference. Here two corrections were tested, the augmented-FSSH (AFSSH) and the energy-based decoherence correction.
A dependence on the employed basis is detected for the AFSSH algorithm, performing better when spin-orbit and external laser field couplings are treated as off-diagonal elements instead of projecting them onto the diagonal of the Hamilton operator.
In the presence of an electric field, the excited state dynamics was found to depend strongly on the vector used to rescale the kinetic energy along after a transition between surfaces.
For SO$_2$, recurrence of the excited wave packet throughout the duration of the applied laser pulse is observed for long laser pulses (>100~fs), resulting in additional interferences not captured by FSSH and only visible in variational multi-configurational Gaussian when utilizing a large amount of gaussian basis functions.
This feature essentially vanishes when going towards larger molecules, such as 2-thiocytosine, where this effect is barely visible in a laser pulse with a full width at half maximum of 200~fs. 
\end{abstract}

\maketitle

\section{Introduction}
Femtosecond time-resolved spectroscopy has progressed drastically throughout the last decades,~\cite{maiuri_ultrafast_2020} challenging the computational excited state dynamics simulations to include explicitly laser pulses.~\cite{mitric_laser_2009,marquetand_nonadiabatic_2011,marquetand_nonadiabatic_2011,bajo_mixed_2012,fiedlschuster_floquet_2016,mignolet_excited-state_2019,makhov_floquet_2018,suchan_importance_2018,penfold_excited_2019,sanz-sanz_field_2019,wohlgemuth_excitation_2020,zhou_nonadiabatic_2020}
Following a laser excitation to some high-lying electronic state, a wave packet can evolve through different potential energy surfaces (PESs), ultimately deactivating to the electronic ground state by radiationless  or radiative processes.
The characterization of these dynamical processes  requires to consider coupled nuclear-electronic motion, reaching beyond the Born-Oppenheimer approximation.
If one also includes the interaction with the laser pulse, the ensuing excited state dynamics will be influenced according to the pulse amplitude and duration, thereby further challenging the calculations as compared to the case of dynamics in the absence of explicit external fields.

Even without explicit laser pulses, the exact quantum treatment of all degrees of freedom in non-adiabatic dynamics is a considerable burden for systems containing more than few atoms.
From the large large number of methods that proliferated,~\cite{gonzalez_book_2020}
the so-called mixed quantum-classical methods employ trajectories as basis functions, moving according to the classical laws of motion to describe the nuclei.
In this category are Ehrenfest,~\cite{parandekar_mixed_2005} $ab$ $initio$ multiple spawning\cite{ben-nun_ab_2000} (AIMS) or fewest-switches surface hopping\cite{tully_molecular_1990,subotnik_understanding_2016} (FSSH) dynamics.
These methods show a favorable scaling with size 
at the price of sacrificing some quantum effects and thus accuracy, in comparison to methods that more readily converge to the exact result like multi-configuratonal time-dependent Hartree\cite{beck_multiconfiguration_2000} (MCTDH), variational multi-configurational Gaussian\cite{worth_novel_2004,lasorne_direct_2007} (vMCG) or full multiple spawning\cite{todd_multi_1996,ben-nun_nonadiabatic_1998} (FMS). 
Contrary to Ehrenfest dynamics and AIMS, FSSH is not directly derived from first principles; instead, it is based on straightforward assumptions that allow a classical trajectory to describe movement and transitions within and between different electronic states.
Although recent investigations employing exact factorization\cite{abedi_exact_2010,agostini_exact_2015,agostini_exact_2018} and the quantum-classical Liouville equation\cite{martens_semiclassical_1997,subotnik_can_2013} have shed light on the nature of FSSH, a proper derivation of FSSH does not exist.
Therefore, systematic improvements of the FSSH algorithm are scarce and known issues, such as the inherent overcoherence,\cite{bittner_quantum_1995,schwartz_quantum_1996} are treated by $ad$ $hoc$ corrections that while remedying the issue at hand in some test cases, fail in others.
The resulting abundance of available options with different parameters to choose from both redeems and curses FSSH:
On the one hand, an appropriate choice of parameters enables at least qualitative accuracy.
On the other hand, most of these parameters are devised on one-- or low--dimensional models, far away from  reality, and when modified in real systems can dramatically affect dynamics.\cite{plasser_strong_2019}

The inclusion of explicit laser pulses in the Hamiltonian is straightforward in wave packet dynamical calculations, but due to the exponential scaling, such simulations are done only in reduced dimensionality. 
Extensions beyond are offered by modified versions of MCTDH,\cite{sanz-sanz_field_2019} vMCG,\cite{penfold_excited_2019} $ab$ $initio$ multiple cloning,\cite{makhov_floquet_2018} or  AIMS,\cite{mignolet_excited-state_2019} showcasing the capabilities of these methods to cope with this additional coupling taking into account many degrees of freedom.
The implementation of explicit laser pulses in FSSH is also possible. \cite{mitric_laser-field-induced_2009,richter_sharc:_2011,marquetand_nonadiabatic_2011}
However, recent comparisons of FSSH to exact quantum results have revealed a strong dependence on the chosen representation of the interaction with the laser pulse. 
Wrong representations were found to fail even for H$_2^+$~\cite{fiedlschuster_floquet_2016,fiedlschuster_surface_2017} where only a set of Floquet states was able to correctly follow the trends of the quantum results.
Similar shortcomings were found for the dissociation dynamics of LiF\cite{mignolet_excited-state_2019} and for single and dual avoided crossing problems employing a set of Floquet states,\cite{zhou_nonadiabatic_2020} showing that no general methodology could be devised so far to treat the coupling correctly.
Overall, it seems that including a laser field into FSSH simulations provides an additional uncertainty as to which parameters are more suitable to use. 

This paper is spurred on bridging recent investigations~\cite{fiedlschuster_surface_2017} that highlighted errors present in FSSH and more apparently in the presence of laser fields and the more pragmatic aim of striving to increase the comparability with experimental data by including laser fields in realistic molecular studies.
With this in mind, we use two model systems, SO$_2$ and 2-thiocytiosine, to test the influence of different FSSH parameters dealing with decoherence, representations, and rescaling options of the kinetic energy after a change between PESs or a frustrated hop.
These tests are conducted both using an explicit laser pulse to excite the system and the much more common approach of neglecting any external field and just placing the wave packet directly in the optical bright state at the beginning of the dynamics.
As a reference, MCTDH (and vMCG) calculations are used to verify the validity of the FSSH simulations, and estimate errors for a given set of parameters, thus showing the grade of applicability of FSSH in realistic molecules.

The remainder of the paper is organized as follows: In chapter~II, the framework of FSSH and the various options available therein are presented. 
Dynamics methods that go beyond FSSH are introduced in chapter~III. 
Chapter~IV describes the nomenclature employed and the methodological details.
Finally, the numerical results are presented in chapter~V followed by the conclusions in chapter~VI.

\section{Fewest-Switches Surface Hopping Background}
Since its original formulation~\cite{tully_molecular_1990} as an improvement to existing surface hopping methodologies, \cite{bjerre_energy_1967, tully_trajectory_1971} FSSH has seen a multitude of adjustments to overcome some of its inherent limitations.\cite{wang_recent_2016,granucci_critical_2007,crespo-otero_recent_2018,subotnik_understanding_2016}
The most severe ones concern energy conservation and the fact that a single independent classical trajectory is unable to describe branching and interference correctly.\cite{schwartz_quantum_1996} 
Instead, the whole wave packet is piggybacking on a single trajectory, resulting in a phenomenon termed overcoherence.
In the following, a short overview of the FSSH methodology with its deficiencies and remedies is given, to provide a backgroud for the parameters that will be used in the subsequent dynamical simulations.


At the core of each FSSH scheme is a classical propagation of the nuclei coupled to a quantum propagation of the electronic wave function bound to the classical trajectory.
In both cases, the propagation is based on a set of electronic states $|\Phi(\mathbf{r},\mathbf{R})_\alpha\rangle$ that are commonly obtained from solving the electronic Schr\"odinger equation.
Surface hopping is not restricted to the adiabatic basis and different sets of states can serve as a basis~\cite{mai_general_2015} (see section below).
For now, a general non-diagonal basis will be assumed although it has been argued that the adiabatic representation is most fitting for the FSSH methodology.\cite{abedi_dynamical_2013}

The electronic wave function ($|\Psi(\mathbf{r},\mathbf{R},t)\rangle$) along each trajectory can then be written as
\begin{equation}
    |\Psi(\mathbf{r},\mathbf{R},t)\rangle = \sum_\alpha c_\alpha(t) |\Phi_\alpha(\mathbf{r},\mathbf{R})\rangle
\end{equation}
where $c_\alpha$ are the coefficients for each electronic state. Their time-dependence is given by
\begin{equation}
    i\hbar \frac{\mathrm{d} c_\beta}{\mathrm{d} t} = \sum_\alpha \left( H_{\beta\alpha}^d(\mathbf{R},t) - i\hbar \frac{\mathrm{d}\mathbf{R}}{\mathrm{d}t} \cdot \mathbf{h}_{\beta\alpha}(\mathbf{R})\cdot \right ) c_\alpha.
\end{equation}
Here, $\mathbf{h}_{\beta\alpha}(\mathbf{R})$ is the non-adiabtic coupling (NAC) vector $\langle\Phi_\beta(\mathbf{r},\mathbf{R})| \frac{\partial\hat{H}_{el}}{\partial \mathbf{R}}| \Phi_\alpha(\mathbf{r},\mathbf{R}) \rangle$, that  
indicates the change of the electronic wave function with variation of the nuclear coordinates.
$H_{\beta\alpha}^d$ is the matrix element of a complete Hamiltonian that contains any arbitrary coupling and can be written as
\begin{equation}
    \label{eq:off-diag}
    H_{\beta\alpha}^d(\mathbf{R},t) = H_{\beta\alpha}(\mathbf{R}) - \hat{\mu}_{\beta\alpha}(\mathbf{R})\mathbf{\varepsilon}(t) + H_{\beta\alpha}^{SOC}(\mathbf{R}).
\end{equation}
The dipole operator $\hat{\mu}_{\beta\alpha}(\mathbf{R})$ mediates coupling between states $\alpha$ and $\beta$ with an external laser field $\mathbf{\varepsilon}(t)$ and $H^{SOC}$ couples states of different multiplicity (here singlet and triplet states) via relativistic spin-orbit coupling (SOC). 
$H_{\beta\alpha}(\mathbf{R})$ is the matrix element of the electronic Hamiltonian in absence of laser and SO couplings. 
In the case of an adiabatic basis, $H_{\beta\alpha}(\mathbf{R})$ equals zero for $\alpha\neq\beta$.

Contrary to the coefficients of the electronic wave function that can be distributed over multiple states at once and will fluctuate across a simulation, the nuclei are restricted to move on only one of the PES in each time step, termed the \textit{active state}.
Non-adiabatic effects are included in FSSH simulations via instantaneous switches between PES when the active state changes.
Multiple algorithms of when the active state should be switched have been proposed --most of them adhering to the concept that the number of switches that occur during a simulation run of a single trajectory should be minimized, thus coining the term of  \textit{fewest-switches}.\cite{tully_molecular_1990,wang_global_2014}
Since a single trajectory is only able to follow one distinct nuclear rearrangement at a time, swarms of trajectories are employed to mimic a nuclear wave packet and obtain meaningful branching ratios or excited state deactivation times.

One of the flaws of FSSH can be readily seen in the use of a set of independent classical trajectories, which prevents the simulation of nuclear quantum phenomena like tunneling or interference.
The advantages of surface hopping, however, rely on its on-the-fly application as a time step in every trajectory only needs to evaluate properties that can be obtained from a single quantum chemistry calculation.
The up-side is that the algorithm itself scales well with the size of the system, only depending on the cost of the corresponding quantum chemistry calculation.
As running FSSH simulations necessitates the calculation of multiple trajectories at once, but the independent trajectory approximation inherent to FSSH allows for all trajectories to be computed independently, it is trivial to parallelize.

\subsection{The Choice of Representation}
\label{sec:rep}

When performing FSSH simulations, different sets of electronic basis states --termed representations-- are available.
While exact wave packet quantum dynamics in a complete basis is invariant to the choice of representation, FSSH is not. 
The nuclei of each trajectory are propagated on the PES of the active electronic state and thus changing the definition of the electronic states will change the PES the nuclei evolve on and in turn the observed dynamics.

The most accessible representation is the so-called molecular Coulomb Hamiltonian (MCH),\cite{richter_sharc:_2011,mai_general_2015,mai_nonadiabatic_2018} in which no coupling between states of different multiplicity and no external field is considered.
Solving the electronic Schr\"odinger equation with the MCH yields a set of diagonal and non-crossing states within each multiplicity.
Additional coupling elements like SOCs or external fields can then be included as off-diagonal elements in the MCH picture.
This MCH set of states can be transformed to a new non-diagonal set of states by a diabatization.\cite{koppel_multimode_2007} 
The PES of diabatic states can be chosen to be smoothly varying and can cross without showing avoided crossings.
For a polyatomic molecule no unique diabatic transformation does exist~\cite{mead_conditions_1982} and the transformation should be chosen as to minimize kinetic-energy coupling and retain a set of chemically relevant states obtained at a reference geometry or to other relevant observable.
A third representation is the one considered in the SHARC approach:~\cite{richter_sharc:_2011,mai_nonadiabatic_2018} here, the complete Hamiltonian $\mathbf{H^d}$ with matrix elements $H_{\alpha\beta}^d$ as in Eq.~\ref{eq:off-diag} is diagonalized.
In this completely diagonal picture (DIAG), the amount of small coupling regions is reduced in favor of more strongly coupled avoided crossings. 
In this work, this DIAG representation will include both, coupling with an external field and SOC, although it is conceivable to include only one of those couplings in the diagonalization while keeping the other as an off-diagonal elements.

A fourth representation that has been employed when it comes to include  external fields is the Floquet representation.\cite{bajo_mixed_2012,chen_proper_2020,zhou_nonadiabatic_2020} 
In the Floquet representation a set of time-independent states is obtained for a continuous wave by diagonalizing the Floquet Hamiltonian.\cite{drese_floquet_1999,korolkov_quantum_2004}
For each state in the original basis a set of infinite new states is created in the Floquet picture that represent the original PES shifted by [-n,-n+1,...,n] times $h\omega$ where $\omega$ corresponds to the frequency of the applied field.
These surfaces are not simply shifted by this amount but show additional avoided crossings at the intersection of states with different photon numbers. 
Performing surface hopping simulations employing a reduced set of Floquet states has been shown to result in surfaces that can coincide with those obtained from exact factorization\cite{fiedlschuster_surface_2017} and thus give the best description for H$_2^+$ in the presence of laser fields --a system where all other representations were found to fail.\cite{fiedlschuster_floquet_2016}
Floquet theory, however, is not guaranteed to give the best FSSH results\cite{zhou_nonadiabatic_2020} and is
only exact in the regime of continuous external fields, and breaking down in the regime of ultra short few-cycle pulses. 
Thus, the Floquet representation will not be applied in the current work.

\subsection{Electronic Decoherence}
\label{sec:deco}

A wave packet traversing a conical intersection branches into a part continuing on the upper state and another propagating on the lower state. 
The two parts can reach different regions of phase space or could interact at a later time.

When it comes to mixed quantum-classical simulation methods, description of a passage through a conical intersection is one of the most decisive steps. 
The FSSH formalism mimics the splitting of the wave packet into two parts while traversing a single conical intersection as a swarm of trajectories that split in two sets following one or the other state.\cite{ibele_molecular_2020} 
Interestingly, the challenge of FSSH to represent the quantum behavior does not lie on the hopping itself, but on the subsequent evolution.
To picture this, we take a look at the evolution of a single trajectory.
When passing through the strong coupling region, the classical trajectory will end up in one of two states, randomly selected based on the coupling strength.
While the classical part of this trajectory is subject to a binary choice, namely which gradient the nuclei will follow, the electronic part tells another story.
The propagation of the electronic coefficients leads to a ratio of state occupations that resembles the branching ratio of the complete wave packet at the conical intersection.
This distribution of electronic occupations is necessary to ensure a binary hopping choice that resembles the real wave packet so that the swarm of trajectories undergoes a reasonable splitting although every trajectory has no information on any other trajectory.
Following the conical intersection, every trajectory is subject to a discrepancy between the classical population (100\% in the currently active state) and the electronic populations, which are a distribution between the two states involved in the strong coupling.
Doing this, the electronic population of a single trajectory mimics both parts of the wave packet at once although the nuclear movement is dictated by the active state and therefore only reminiscent of this single branch of the wave packet.
Dragging this "wrong" part of the wave packet along is termed \textit{overcoherence}. 
When another coupling region is encountered, there is interaction both with the "right" and the "wrong" parts of the wave packet and wrong hopping probabilities will be predicted as the presence of this second part of the wave packet is nonphysical.

In the last decades, a plethora of modifications (decoherence methods) to the plain surface hopping have been presented to remedy overcoherence, i.e. trying to adapt the electronic populations in a more or less physical way to resemble the quantum dynamical results.
In this work, we will work with two decoherence methods: the energy-based decoherence correction (EDC) \cite{granucci_critical_2007} and the augmented fewest switches surface-hopping (AFSSH).\cite{subotnik_new_2011,jain_efficient_2016}
EDC is based on the simple assumption that the branched part of the wave packet in another state will dephase and move into another spatial part where the interactions between these two wave packet parts vanish.
For a single trajectory this means that electronic population in non-active states should slowly decay because any branched part of the wave packet in another state will dephase and move into another spatial part where the interactions between these two wave packet parts vanish.
This decay is realized by modifying the electronic populations\cite{granucci_including_2010} ($p_i$) of every non-active state in every time step via
\begin{equation} 
  p_i'=p_i\cdot exp \left(- \frac{|E_i-E_\alpha|}{\hbar}\frac{E_{kin}}{E_{kin}+ C}\right)
\end{equation}
where $E_i$ and $E_\alpha$ are the energy of the $i$th non-active and the currently active state, respectively. 
$E_{kin}$ is the kinetic energy and $C$ is a parameter commonly set~\cite{zhu_coherent_2004} to 0.1~$E_h$. 
Any population that is reduced from non-active states is added to the population of the active state to keep the overall population constant. 
The modified populations ($p_i'$) are then used subsequently.
The AFSSH mechanism\cite{jain_efficient_2016} is more intricate as it tries to track where any part of the wave packet in a non-active state is moving to.
In every time step, gradients in the non-active states are collected and auxiliary trajectories propagated in those states. 
If the trajectories deviate too far or too fast from the active trajectory, the population in this non-active state is considered dephased and set to 0.

Both EDC and AFSSH decoherence corrections have been shown to be improve dynamics over the case of using no decoherence correction at all in a set of test cases without including laser fields.\cite{granucci_critical_2007,subotnik_new_2011,jain_efficient_2016} 
The EDC has been applied to the excited state dynamics of LiF in the presence of different laser pulses, where no significant improvement over the basic FSSH algorithm without any treatment of the overcoherence was observed when compared to exact quantum simulations.\cite{mignolet_excited-state_2019}
Yet, one should keep in mind that the presented decoherence corrections only explicitly tackle the problem of overcoherence but do not improve FSSH results when it comes to a wave packet recombination event.
These recombination events can be important in very specific systems but have minor influence in most cases\cite{subotnik_understanding_2016}.

\subsection{Energy Conservation}
\label{sec:Econ}

Propagating a swarm of non-interacting trajectories raises an important issue when it comes to energy conservation throughout the dynamics.
The total energy contained within the system should not change in the course of the dynamics.
This energy can be distributed into kinetic and potential energy in all nuclear and electronic degrees of freedom.
When a hop between PES occurs, the potential energy of the trajectory undergoes an instantaneous change because the active state is switched. 
This behavior would cause discontinuities in the total energy of the single trajectory and possibly even that of the total ensemble.
To ensure energy conservation of the total ensemble one typically enforces total energy conservation along each individual trajectory.
For this, every change in the potential energy of the trajectory induced by a surface hop is compensated by adapting the kinetic energy correspondingly.
In Tully's original prescription,\cite{tully_molecular_1990} only the nuclear momenta parallel to the  NAC vector  $\mathbf{h}_{\alpha\beta}$ are rescaled.
If NAC vectors are not available in a simulation, the gradient difference vector $\mathbf{g}$ can be used as a substitute.
\begin{equation}
    \mathbf{g}_{\alpha\beta} = \mathbf{F_\beta} - \mathbf{F}_\alpha
\end{equation}
with $\mathbf{F_\alpha}$ and $\mathbf{F_\beta}$ being the gradient in the active state and the non-active state, respectively.
Even simpler is to rescale the full nuclear momentum,  which offers an intriguing simplicity as no NACs vectors need then to be computed. 
Hops to a lower PES result in an increase in kinetic energy while transitions to higher-lying PESs are accompanied by a reduction in the kinetic energy of the system.
When trying to enforce any of the mentioned energy conservation procedures, a special case can be encountered if the FSSH algorithm wants to switch the active state to a higher energy PES but the kinetic energy in the system is insufficient to bridge the energy gap to this PES.
Such an attempt at hopping is rejected by most algorithms and fittingly coined as a \textit{frustrated hop}.
The presence of frustrated hops has been noted early on and it was found that they are necessary to retain detailed balance within surface hopping simulations.\cite{schmidt_mixed_2008,carof_detailed_2017}
However, the number of observed frustrated hops can vary significantly, depending on which of the rescaling schemes is employed.
This is due to the different amounts of available kinetic energy in e.g. rescaling along $\mathbf{h}$ or rescaling along the full velocity vector.
More energy is available in the full velocity vector as the velocity component along the NAC vector is included therein.
Using the full kinetic energy rescaling scheme, i.e. enlarging a system, e.g. by including additional non-interacting molecules at large distances, increases the total kinetic energy thus enabling the possibility to jump to higher-lying states than when  non-interacting molecules are absent.
If rescaling along $\mathbf{h}$ is used, inclusion of additional non- or weakly interacting molecules has none or limited influence on the NAC vector and thus the same amount of energy as without the additional molecules is available for hopping to surfaces of higher energy.
Rescaling along the NAC vectors has been found to give results that are in best agreement with quantum dynamics and is then treated as the preferred option when NAC vectors are available.\cite{carof_detailed_2017,plasser_strong_2019}

Unfortunately, the concept of strict energy conservation throughout the dynamics does not hold in the presence of an external field, which stimulates absorption or emission of photons, adding or subtracting energy to the system.
When an external field is applied, two options are available to deal with this flow of energy:
One is to suspend the conservation of the total energy for the duration of the pulse.
Then, the nuclear momenta will not be adapted at a surface hopping event and the trajectory will be propagated on a different PES using the same nuclear momenta.
Therefore, any changes in the potential energy are also found in the total energy.
Another option is to find criteria to verify whether a hop is laser-induced.
This can be realized by tracking the change in the electronic coefficients due to the laser coupling as opposed to changes in the coefficients due to NACs.\cite{bajo_interplay_2014} 
Alternatively, one can set an energy interval around the center frequency of the external field\cite{richter_sharc:_2011} and if the energy gap at a hopping event falls within this interval, the hop is labelled field-induced and thus no rescaling of the velocities is performed.

\section{Reference methods}

\subsection{Multi-configurational time-dependent Hartree}
Introduced in 1990, the MCTDH\cite{meyer_multi_1990,beck_multiconfiguration_2000,worth_using_2008} method is one of the most versatile methods to simulate non-adiabatic processes, due to the inherent possibility to converge towards the exact solution.
Briefly, MCTDH is based on an expansion of the wave function using a time-dependent basis of single-particle functions (SPFs), $|\varphi\rangle$:
\begin{eqnarray}
    |\Psi(q_1,...q_f,t)\rangle=&&\sum_{j_1=1}^{n_1} \dots \sum_{j_p=1}^{n_p} A_{j_1\dotsc j_p}(t)|\varphi_{j_1}^{(1)}(Q_1,t)\dots \varphi_{j_p}^{(p)}(Q_p,t)\rangle \nonumber \\ 
    =&&\sum_J A_J\Phi_J
\end{eqnarray}
where the coordinates of $\Psi$ are $f$ explicit degrees of freedom ($q_f$), which can be combined to form a set of actual coordinates $Q_i$. 
Each $Q_i$ then represents one or more degrees of freedom at once, in a process called "mode combination".
$\Phi_J$ is a single Hartree-product preceded by its corresponding coefficient $A_J$.
Through the Dirac-Frenkel variational principle, the best suited expansion coefficients and SPFs can be determined directly for each time-step.
Therefore, MCTDH can be understood as a method that allows describing the evolving wave packet in a reduced number of used basis functions for each time-step, and thereby minimizing the computational effort while maintaining a maximum amount of accuracy.

\subsection{Variational multiconfigurational Gaussian}

A drawback of MCTDH is that the SPFs themselves still use a static grid and the whole method therefore relies on a functional form of the PES.
Subsequent, alternative formulations have been proposed resulting in the development of the vMCG\cite{worth_novel_2004,lasorne_direct_2007} method, where frozen Gaussians are used as basis functions instead of SPFs.

In short, the wave function ansatz for $\Psi$ then reads:
\begin{equation}
    |\Psi(\mathbf{r},\mathbf{R},t)\rangle=\sum_i A_i^{(s)}(t)|\varphi^{(\alpha)}(\mathbf{r};\mathbf{R})\chi_j^{(\alpha)}(\mathbf{R},t)\rangle
\end{equation}
where the electronic wave function for a state $\alpha$, $|\varphi^{(\alpha)}(\mathbf{r};\mathbf{R})\rangle$ is multiplied by a set of Gaussian basis functions (GBFs) $|\chi_j^{(\alpha)}\rangle$ and time-dependent expansion coefficients $A_j^{(s)}$.
As in MCTDH, the equations of motion can be derived variationally and result in the propagation of both the expansion coefficients and the parameters of the GBFs.
The coupled motion of both the coefficients and the GBF parameters results in a "quantum" movement that goes beyond simple classical motion.
Also similar to MCTDH, in the limit of infinite basis functions (SPFs for MCTDH and GBFs in vMCG), the complete space is covered in basis functions and the exact dynamics is obtained.
The vMCG algorithm has the advantage that it can be employed in an on-the-fly fashion with no need to rely on pre-computed PESs.\cite{worth_solving_2008}

\section{Computational Details}

\subsection{Nomenclature}
\label{subsec:names}
In order to distinguish between the different combinations of FSSH parameters employed in this work, the following short-hand notation will be used:
\begin{equation}
    ^{\mathrm{REPRESENTATION}}\mathrm{DECOHERENCE}_{\mathrm{rescaling}}^{\mathrm{frustrate~hops}}
\end{equation}

The possible options for each of the keywords are listed in Table~\ref{tab:keywords} and explained as follows.\\
(i) Representation. Two representations for the propagation are used, the MCH or the completely diagonal picture (DIAG) previously introduced in Section~\ref{sec:rep}\\ 
(ii) Decoherence. To correct for overcoherence we use AFSSH or EDC, see Section~\ref{sec:deco}.
Additionally, the NONE option indicates that no decoherence correction is included.
\\
(iii) Rescaling after a surface hop. 
For the treatment of velocity rescaling after a transition between states, we rescale the velocity vector after a hopping event along the full velocity vector ({v}), the non-adiabatic coupling vector of the involved states ({h}) or the gradient difference vector ({g}), or we do not rescale the momenta at all ({none}).\\
(iv) Frustrated hops. If an insufficient amount of kinetic energy is available to compensate for a hop to a higher-lying PES during the rescaling process, the hop is cancelled and termed frustrated.
At such frustrated hops, the velocities of the trajectory can be reflected.
The same set of vectors as in (iii) is available to reflect the momenta, resulting in the options -{v}, -{h} and -{g}, where the minus indicates the reflection event. 
Alternatively, the velocities can be kept at a frustrated hop without any reflection, which is labelled "+".\\

\begin{table}
\begin{tabular}{cccc}
\hline
  REPRESENTATION & DECOHERENCE & rescaling & frustrated hops\\
\hline
 
  MCH            &  AFSSH       & {v}      & -{v}              \\
  DIAG           &  EDC        & {h}         & -{h}             \\
                 &  NONE        & {g}         & -{g}             \\
                 &                 & none         & +             \\
\hline
\end{tabular}
\caption{FSSH options employed for dynamics.}
\label{tab:keywords}
\end{table}

Using this notation, a FSSH setup in the diagonal representation, using the AFFSH decoherence correction and rescaling along the NAC vectors at hopping events and no reflection at frustrated hops would be denoted as $^{\mathrm{DIAG}}AFSSH_{h}^+$.
Note that not all combinations of these parameters result in stable setups for FSSH dynamics: e.g. in the MCH representation, the non-adiabatic coupling vector $\mathbf{h}$ between singlet and triplet states is zero and therefore cannot be used to rescale in simulations including singlet and triplet states.
Apart from these incompatibilities, all possible combinations between options (i), (ii) and (iii) have been used.
Throughout this work, a consistent pairing of the used vectors in (iii) and (iv) has been used:
This means for example, that when using the gradient difference vector in (iii), reflection of frustrated hops in (iv) cannot be conducted along -{v} or -{h} but has to occur along -{g}.
Using the option to not rescale the kinetic energy at all in (iii) leads to zero frustrated hops and is therefore only paired with "+" for the treatment of frustrated hops.
The option "+" in (iv), however, can be paired with all rescaling vectors (iii) to shed light on the difference between dynamics where reflection takes place at frustrated hops compared to dynamics, where the same trajectory is not reflected at this point.

\subsection{Error quantification}

To quantify the difference between a reference MCTDH calculation and a FSSH calculation with a particular set of parameters, we define an error $\epsilon$ calculated as  
\begin{equation}
    \epsilon = \frac{max_t}{\Delta t}\sum_t^{max_t} \sum_{i=1}^{n_{states}} |p_{i,t}-p_{i,t,ref}|
    \label{eq:error}
\end{equation}
where $max_t$ is the final simulated time, $\Delta t$ is the employed time step, $p_i,t$ and $p_{i,t,ref}$ are the populations of state $i$ taken from the investigated and the reference dynamics at time $t$. 
If the populations of the reference and the investigated dynamics are identical for all time steps, $\epsilon$ equals zero.
The maximum value of this error is achieved only if for every time step of the dynamics, all population in the investigated dynamics is found in states which are not populated at all in the reference dynamics.
For example, if all population in the reference dynamics is in state $a$ while the investigated dynamics always run in state $b$, the finale error equates to $|p_{a}-p_{a,ref}|+|p_b-p_{b,ref}|+\sum_{i\neq a,b}^{n_states-2}|0-0|=2$. 
The error is further split into contributions from the singlet and triplet states, by summing the errors over all singlet ($\epsilon_{sing}$) and all triplet states ($\epsilon_{trip}$), respectively.

The error measure in Eq.~\ref{eq:error} is not well suited to describe the error in the dynamics if a laser pulse is present:
As mentioned above, the maximum error is 2, achieved if all the population is found in differing states for both the reference and the investigated dynamics.
However, when we imagine an extreme case, where a weak laser excites only 1\% of the ground state population to higher-lying states in both the reference and the investigated dynamics, one would obtain a maximum error of 0.02 due to the agreement in the 99\% of population staying correctly in the S$_0$ ground state.
Therefore, we use two differently calculated errors for simulations that include laser fields, in order to provide errors that can be compared to errors from dynamics without a laser field:
First, the deviation in the ground state population ($\epsilon_{S_0}$) is calculated according to Eq.~\ref{eq:error} only using the ground state population; this will indicate the capability of a given set of FSSH parameters to describe the initial excitation process.
To factor out the differences in the excitation process itself from the subsequent excited state dynamics, the error in the excited state populations ($\epsilon^r$) is calculated using a renormalized excited state population.
This is done by renormalizing the population in the excited states in every time step to 1 for both the reference and the FSSH dynamics yielding $\epsilon^r$ (the superscript thus indicates the renormalization within this error measure:
\begin{equation}
    \epsilon^r = \frac{max_t}{\Delta t}\sum_t^{max_t} \sum_{i=2}^{n_{states}-1} \left | \frac{p_i,t}{1-p_{S_0,t}}-\frac{p_{i,t,ref}}{{1-p_{S_0,t,ref}}}\right |
    \label{eq:renorm_error}
\end{equation}

To avoid numerical instabilities where the denominator in $\frac{p_i,t}{1-p_{S_0,t}}$ or $\frac{p_{i,t,ref}}{{1-p_{S_0,t,ref}}}$ is close to zero, $\epsilon^r$ is only calculated for time steps where the S$_0$ populations in both dynamics are below 0.98.

To compare the diabatic MCTDH (and vMCG populations) to FSSH populations and to obtain diabatic populations for FSSH dynamics that contain both information about the distribution of active surfaces and the corresponding electronic populations at the same time, the methodology applied in Ref~\citenum{landry_communication_2013} has been employed. 

\subsection{Dynamical Propagations}

The FSSH simulations have been carried out with the SHARC\cite{mai_general_2015,mai_sharc2.0:_2018} program suite using pySHARC\cite{plasser_highly_2019} to drastically reduce I/O overhead.
A set of 1000 nuclear initial conditions obtained from a ground state Wigner sampling\cite{wigner_quantum_1932} is used for both SO$_2$ and 2-thiocytosine.
The simulations employed a time step of 0.5~fs for the nuclear propagation in the field-free case and a nuclear time step of 0.05~fs in the presence of an external field. 
The electronic wavefunction is propagated in a locally diabatic basis with nuclear time steps of 0.02~fs and 0.002~fs in the field-free and field-including case, respectively.\cite{granucci_direct_2001} The reduction of step size when including an electromagnetic field is necessary to capture the rapidly oscillating field.

As explained above, various FSSH options and modifications will be tested and compared to reference quantum dynamical results.
Note however, that the representation in which FSSH is performed is never the purely diabatic representation of the LVC model (see below).
Two different sets of initial electronic coefficients have been employed:
For the simulations without an explicit laser field, the initially active electronic state at the start of the FSSH dynamics is set to the MCH state that has the largest overlap with the diabatic bright state of SO$_2$ and 2-thiocytosine for each initial condition.
After setting the initial electronic state, the electronic coefficients are adapted correspondingly so that the initial electronic population at the start of the dynamics amounts to 1 for the diabatic bright state.
This methodology represents an instantaneous excitation of the complete ground state wave packet to a single diabatic bright state, as it would follow an ideal instantaneous $\delta$-pulse.
For simulations in the presence of a laser field, the initial electronic state was set to the lowest energy state and no modification of the initial electronic coefficients was done.

The QUANTICS package has been employed to run MCTDH and vMCG dynamics.\cite{Quantics_program}
In MCTDH, the Adams-Bashforth-Molton  predictor-corrector  integrator of 6th order and the multi-set formalism have been used.
Convergence of MCTDH dynamics was deemed to be reached if both of the following criteria have been met:
First, the weight of the last SPF assigned to a degree of freedom did not exceed a value of 0.001.
Second, the number of grid points in a mode was taken to be sufficient if it was larger than $\langle n \rangle+3\cdot\langle dn \rangle$ for all states.

vMCG calculations were run in a single-set formalism which was found to be computationally more efficient than the multi-set formalism and converged faster to the MCTDH results as well, which is in line with previous observations.\cite{penfold_excited_2019} 
Integrals between gaussian wave packets have been calculated up to 4th order.
The Runge-Kutta integrator of fifth order has been used. 
The number of considered gaussian basis functions in vMCG ranged from 10 to 100.
For both, vMCG and MCTDH, initial conditions were obtained by populating the vibrational ground state of the lowest-energy electronic state of all normal modes considered.
In simulations without an external field, this initial wave packet was set to start in the diabatic bright state while for simulations in the presence of an external field, no additional steps were taken.
The input and operator files used can be found in the supporting information.
 
All FSSH, MCTDH, and vMCG simulations were run for 400~fs.

\subsection{Definition of the laser pulse}

The coupling with the external field 
is described within the the semi-classical dipole approximation and can be written as 
$-\hat{\mu}_{\beta\alpha}(\mathbf{R})\mathbf{\varepsilon}(t)$
neglecting any further interaction terms.
The pulse has a Gaussian shape, defined as 
\begin{equation}
\label{eq:pulse}
    \mathbf{\varepsilon}(t) = \textbf{e}_i\varepsilon^0_{t_p} cos(\omega_i(t-t_0)+\eta) exp\left[-4\mathrm{ln}2\left(\frac{t-t_0}{t_p}\right)^2\right],
\end{equation}
with field amplitude $\varepsilon^0_{t_p}$, carrier frequency $\omega_i$, carrier envelop phase $\eta$ and pulse duration $t_p$ equivalent to the full width at half-maximum (FWHM). 
We assume a linearly polarized pulse along the x-direction ($\textbf{e}_i=x$) of the transition dipole moment and a phase $\eta$=0.
The frequency is set to be in resonance with the brightest state of the the corresponding molecules: 4.49~eV for SO$_2$ and 3.92~eV for 2-thiocytosine.
Seven different values for $t_p$ are used with the corresponding centers of the pulse ($t_0$) in parentheses: 2~fs ($t_0=10$~fs), 10~fs ($t_0=30$~fs), 17~fs ($t_0=40$~fs), 30~fs ($t_0=70$~fs), 50~fs  ($t_0=90$~fs), 100~fs ($t_0=140$~fs), and 200~fs  ($t_0=200$~fs).

The field amplitude of the laser field for a given $t_p$, $\varepsilon_{t_p}^{0}$, was varied for different lengths of the laser pulse according to $\varepsilon^0_{t_p}=\varepsilon^0_{17}\cdot \sqrt{\frac{17}{t_p}}$.
This way the pulse energy (area of intensity throughout the pulse duration) for different laser pulse lengths is kept constant.
The pulses with a $t_p$ of 17~fs serve as reference pulses for the determination of the field amplitudes for other pulse lengths and where set to $\varepsilon^0_{17}=0.03$~a.u. (15.44~GV/m) for SO$_2$ and $\varepsilon^0_{17}=0.01$~a.u. (5.15~GV/m) for 2-thiocytosine.
Using these amplitudes, about half of the S$_0$ population in the FSSH simulations was excited in both molecules, which was deemed a good tradeoff between minimizing non-linear behavior and Rabi-oscillations while still exciting enough population in FSSH to yield reasonable statistics.


\subsection{LVC model}
The PES employed for the dynamical simulations are parameterized using a  linear vibronic coupling (LVC) model.\cite{worth_beyond_2004}
Vibronic coupling models are diabatic representations of the electronic states close to a reference point and capable of describing dynamics and conical intersections close to the reference point. 
A suitable reference point capable of describing interactions with all excited states directly after initial photoexcitation is the ground state equilibrium geometry.
The spatial dependence of the excited states is then cast into mass-frequency scaled normal mode coordinates using the normal modes of the ground state as a basis.
When truncating this Taylor expansion at first-order, only linear terms  that depend on a single normal mode displacement $Q_i$ are obtained, resulting in
\begin{equation}
    \mathbf{H}_{LVC} = \mathbf{H}^{(0)} + \mathbf{W}^{(1)}
\end{equation}
where $\mathbf{H}_{LVC}$ is the diabatic LVC Hamiltonian and $\mathbf{H}^{(0)}$ contains the zero-order harmonic potential approximations to the PES 
\begin{equation}
    \mathbf{H}^{(0)} = V_0 \mathbf{1} , V_0 = \sum_i \frac{\hbar\omega}{2}Q_i^2.
\end{equation}
Here, $V_0$ is the harmonic ground state potential along every normal mode $i$.
The first order terms in $\mathbf{W}^{(1)}$ are state-specific and consist of electronic energy shifts $\epsilon$, intrastate gradients $\kappa$ and couplings between two states $\lambda$:
\begin{equation}
    W_{nn}^{(1)} = \epsilon_n + \sum_i \kappa^{(n)}_iQ_i
\end{equation}
\begin{equation}
    W_{mn}^{(1)} = \sum_i \lambda^{(mn)}_iQ_i
\end{equation}    

Truncating the Hamiltonian after the first-order terms results in a rather crude PES able to accurately capture the form of potential only near the reference point. 
Parameters for the systems investigated here were taken from Ref~\citenum{plasser_highly_2019}, that also demonstrates that the LVC approximation is able to reproduce the main characteristics of the corresponding on-the-fly dynamics. 

For use in the remainder of the work, the diabatic states and properties were transformed to either the MCH or completely diagonal representation during the propagation.

\section{Results and discussion}

\subsection{SO$_2$}
The excited state dynamics of SO$_2$ after irradiation has been extensively investigated in the last decade, illustrating a complex interplay between singlet and triplet states.\cite{Leveque_theoretical_2014, Leveque_excited_2014, mai_non_2014} 
In this work, we use a LVC Hamiltonian that contains 4 singlet and 3 triplet states obtained with multi-reference configuration interaction including single excitations, which is able to reproduce the main features of the full-dimensional excited state dynamics.\cite{plasser_highly_2019} 
In order to find an optimal set of surface hopping parameters able to describe the excited states dynamics of SO$_2$ \textit{in the presence} of an explicit laser pulse, we first validate different parameter sets for the ensuing excited state dynamics \textit{in the absence} of the laser field. 

\subsubsection{SO$_2$ dynamics in the absence of a laser field}
\label{subsec:so2_nolaser}

\begin{figure}[ht]
\includegraphics{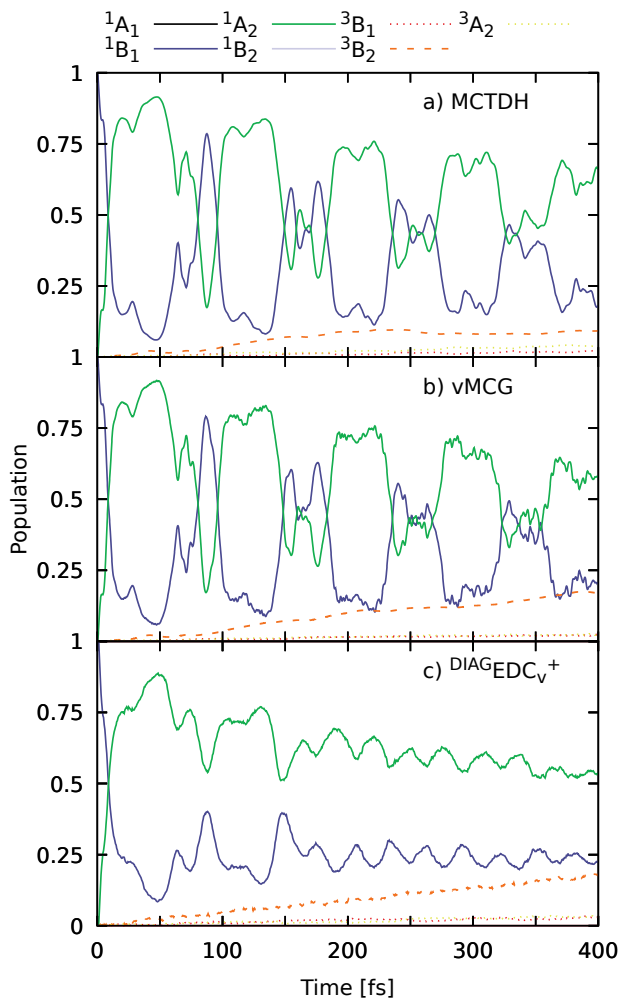}
\caption{\label{fig:so2_nolaser} Time-resolved population of the electronic states involved in the excitation of SO$_2$ starting from the diabatic $^1$B$_1$ state using different methods: (a) MCTDH, (b) vMCG dynamics using 75GBFs  and (c) FSSH with the parameter set $^{\mathrm{DIAG}}$EDC$_{v}^+$. Singlet states are depicted by solid lines, triplet states in dashed lines.}
\end{figure}

To simulate the non-adiabatic dynamics of SO$_2$ without an explicit laser pulse excitation, 
the ground state nuclear wave packet is vertically placed in the bright $^1$B$_1$ state. 
Figure~\ref{fig:so2_nolaser} shows the results obtained with MCTDH, vMCG and FSSH dynamics.
The MCTDH dynamics (Fig.~\ref{fig:so2_nolaser}a), which will serve as a reference throughout, shows ultrafast population transfer from the initially populated $^1$B$_1$ state to the $^1$A$_2$ state.
Within the first  50~fs, there is almost complete depletion of the $^1$B$_1$ population, which then oscillates in par with that of the $^1$A$_2$ during the rest of the propagation.
This marked oscillatory behavior is due to the closeness of the respective minima, both in energy and phase space.
Therefore, only limited stabilization of one diabatic state with respect to the other can take place, resulting in the repeating pattern that gets more complex over time.
During the first 400 fs, about 10\% of the population crosses to the triplet manifold, where the $^3$B$_2$ state is populated almost exclusively --in line with the also MCTDH simulations performed on ab initio potentials by L\'ev\^eque $et$ $al$.\cite{Leveque_theoretical_2014} 

The vMCG dynamics using 75~GBFs (Fig.~\ref{fig:so2_nolaser}b) show only small differences compared to the MCTDH reference.
The excellent agreement of the vMCG singlet populations with the MCTDH ones illustrates the strength of the method.
The oscillating behavior between the $^1$B$_1$ and the $^1$A$_2$ states is very well reproduced, with minor deviations at later times. 
The differences in the triplet states, however, are more pronounced. 
There is a continuous transfer to the triplet states, mostly to the $^3$B$_2$ state, which is not observed in the MCTDH dynamics, in which the population in this state stagnates after 250~fs.
To quantify the differences between vMCG and MCTDH dynamics, we calculate the error $\epsilon$ following Eq.~\ref{eq:error}. 
This results in the average deviation between the reference populations and the dynamics in question in each time step.
The so-derived error is $\epsilon$=0.069 and it is attributed almost equally to the $^1$B$_1$, the $^1$A$_2$, and the $^3$B$_2$ states.
Increasing the number of used GBFs is expected to decrease the error until convergence towards the exact result (see Section~S1~C in the supporting information for the $\epsilon$ values for simulations using different numbers of GBFs).

The FSSH simulations yield different results depending on the choice of the parameters discussed above (representation, decoherence correction, rescaling of the kinetic energy after a hopping event and reflection of the kinetic energy after a frustrated hop).
For the sake of briefness, we show in  Fig.~\ref{fig:so2_nolaser}c
only the populations for a single parameter set: $^{\mathrm{DIAG}}$EDC$_V^+$.
Compared to the MCTDH reference, the FSSH dynamics start with a similar transfer between the $^1$B$_1$ and the $^1$A$_2$ state, matching the time needed for this transfer.
However, after the initial 50~fs the oscillatory transfer between both states is far less pronounced than it was in MCTDH.
The damped oscillations in FSSH follow the general trend of the MCTDH populations, with a slow decline in $^1$A$_2$ population at later times.
The triplet state population is reminiscent of the vMCG dynamics, with a continuous transfer to the $^3$B$_2$ state.
The error associated to this simulation gives $\epsilon$=0.248, being the the main sources of error the damped oscillations and the transfer to the $^3$B$_2$ state.
Unsurprisingly, FSSH has a larger error than vMCG (approximately 3 times larger).

\begin{figure*}
\includegraphics{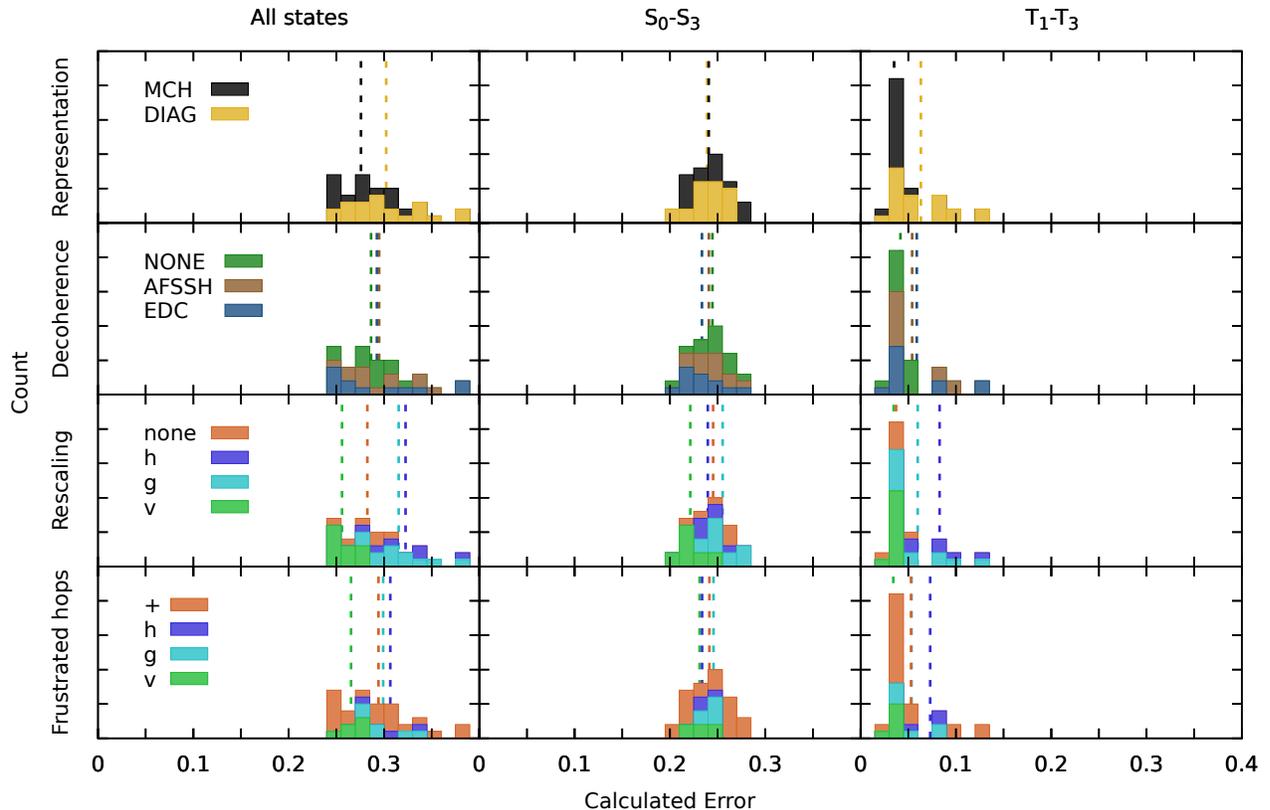}
\caption{\label{fig:so2_nolaser_errors} Calculated $\epsilon$ values for each set of surface hopping parameters employed for the simulation of SO$_2$ represented as a stacked histogram combining errors that fall into bins of size 0.015. The contribution of each parameter towards the histogram is colored correspondingly. The dashed lines represent the average error associated with the identical colored parameter for the given panel. Each row displays the contribution of all parameters associated with a specific setting, as indicated in the label. The left panel shows the combined error ($\epsilon$) running over all states while the middle and right panels display the calculated error over singlet ($\epsilon_{sing}$) and triplet states ($\epsilon_{trip}$), respectively.}
\end{figure*}

In order to investigate the influence of different FSSH parameters 
all sensible combinations of parameters according to Section~\ref{subsec:names} have been used to run FSSH dynamics.
Figure ~\ref{fig:so2_nolaser_errors} collects all the deviations against the MCTDH reference for each set. 
Find a list of all obtained $\epsilon$ values in Tables~SI-SIII in the supporing information.

The $\epsilon$ ranges from 0.240 to 0.383, meaning that an unsuitable  combination of surface hopping parameters can increase the error with respect to the best possible set by 50\% .
Each column of Fig.~\ref{fig:so2_nolaser_errors} gives the stacked error summed over all the states, singlet and triplet states, respectively,  while each row depicts the influence of a particular option in the FSSH algorithm.

When comparing the influence of the representation on the errors, it is found that using the MCH representation gives smaller deviations than its completely diagonal counterpart.
While this does not hold for $\epsilon_{sing}$, where the average errors obtained using either representation are almost identical, big differences can be observed in $\epsilon_{trip}$: all MCH simulations yield an almost identical error while only half of the simulations employing the diagonal representation yield comparable errors.
The difference observed in the average error due to the representation is therefore due to the better performance of the MCH representation to describe the triplet state populations.
The decoherence correction inflicts very small differences among the three different parameters investigated.
However, the FSSH dynamics on the investigated SO$_2$ model system is very sensitive to the choice of the vector along which rescaling of the kinetic energy should be conducted.
Here, two clear favorites emerge of which rescaling along the total velocity vector outperforms the crude option of not changing the velocities at all after switching the active state.
Yet, both of these parameters are found with significantly smaller average error than rescaling along the non-adiabatic coupling vector (\textbf{h}) or the gradient difference vector (\textbf{g}), which gives the highest average error of all investigated options for all parameters.
The bad performance of the \textbf{h} and \textbf{g} parameters is mainly due to the triplet states, where they are the cause for almost all large deviations from the MCTDH reference.
Interesting to note here, is that choosing to rescale along \textbf{h} or \textbf{g} significantly reduces the amount of available kinetic energy that can be used to reach a higher-lying surface, as compared to option of rescale along the complete velocity vector and of course the option of not to adapt the energies at all.
When comparing the number of successful and frustrated hops for two sets of parameters that differ only in the rescaling vector, e.g. $\mathbf{v}$ and $\mathbf{g}$, a total of 3629 executed- versus 3833 frustrated hops is found for $^\mathrm{MCH}$EDC$_g^+$ while 7484 hops against only 1041 frustrated hops are found for $^\mathrm{MCH}$EDC$_v^+$ on the complete set of 1000 trajectories.
The higher available kinetic energy in the complete kinetic energy vector therefore enables a lot of jumps that are forbidden when just using the gradient difference vector for rescaling.
We noted that switching to the diagonal basis increases the total number of observed hops drastically (39908 for $^\mathrm{DIAG}$EDC$_g^+$ and 41738 for $^\mathrm{MCH}$EDC$_v^+$) due to the presence of more trivial and avoided crossings, as the triplet states are split into their respective components.
Although the total number of hops increases in the diagonal picture, the amount of frustrated hops stays on a comparable level, with 4664 frustrated hops ($^\mathrm{DIAG}$EDC$_g^+$) and 859 ($^\mathrm{DIAG}$EDC$_v^+$).
This might shed some light on why the  largest errors are associated with using either of the two decoherence corrections in the diagonal representation along with rescaling the kinetic energy after a hop along \textbf{h} or \textbf{g}, as an increased number of hops slows down both types of decoherence corrections.
In general, for the treatment of frustrated hops, reflection of the full velocity vector results in the lowest average error.
As this option is always tied to using the full velocity vector for estimating if sufficient kinetic energy is present, which is the largest amount of available energy apart from not rescaling the energy at all, the number of frustrated hops is very low in these dynamics.
The notion of "few frustrated hops equals better agreement between populations" is reinforced when realizing that most best-performing sets did not reflect frustrated hops at all.
However, combinations of rescaling along \textbf{h} or \textbf{g} with a complete neglect of frustrated hops gave the highest disagreement for singlet and triplet populations, showcasing that a correct treatment of frustrated hops is needed in these cases.

The findings can be summarized as follows: 
(i) The MCH picture results in a stable description of the triplet population. 
(ii) No real favorite  decoherence correction emerges. 
(iii) Rescaling along one of the physically more sound options \textbf{h} or \textbf{g} results in larger deviations from the MCTDH dynamics, especially when paired with the parameter of not reflecting frustrated hops. 
Good results were achieved when using \textbf{v} for rescaling the kinetic energy combined with a continuation along the current velocity when encountering frustrated hops.
(iv) A very bad combination of parameters  for the description of the triplet states emerges in the form of using a decoherence correction in the diagonal representation and rescaling along  \textbf{h} or \textbf{g} using non-reflected frustrated hops.

\subsubsection{SO$_2$ dynamics in the presence of a laser field}
\label{subsec:so2_laser}

\begin{figure}[ht]
\includegraphics{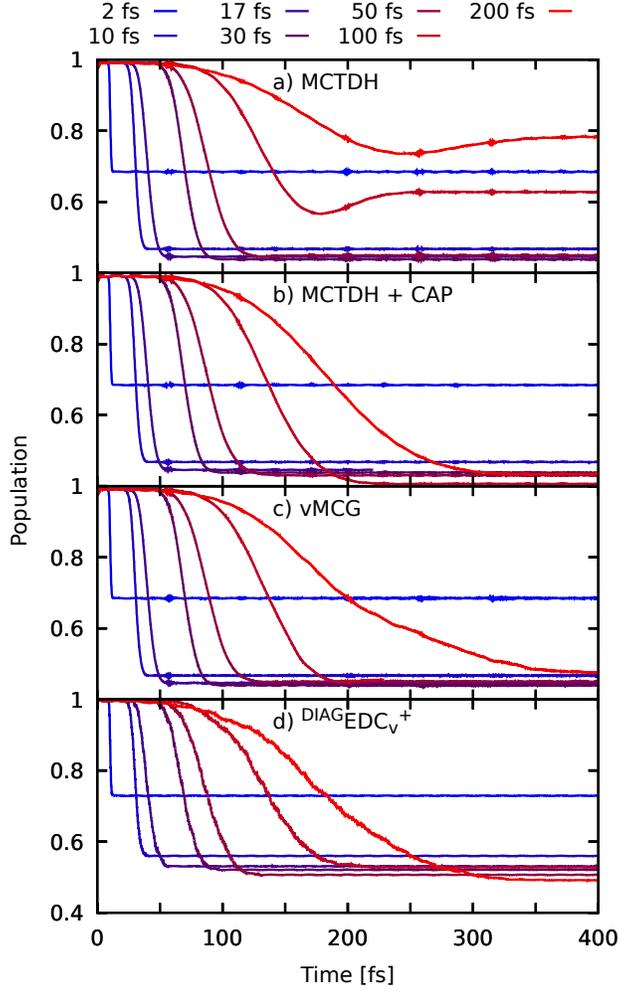}
\caption{\label{fig:so2_s0} Time-resolved S$_0$ population for 7 different sets of dynamics, employing laser pulses differing in the $t_p$ of the envelope function and the field amplitude $\varepsilon^0_{t_p}$ 
(see Eq.~\ref{eq:pulse}). Panel a) displays the MCTDH dynamics, while panel b) and c) show surface hopping dynamics with the parameter set $^{\mathrm{DIAG}}$EDC$_{v}^+$ and vMCG dynamics using 75~GBFs.}
\end{figure}

In the presence of a laser field, both the excitation and the subsequent excited state evolution of the population is influenced by the laser pulse. 
We shall investigate both processes separately, starting with the excitation process itself.
The effect of applying longer pulses to excite population from the $^1$A$_1$ ground state is shown in Fig.~\ref{fig:so2_s0}. 
Before discussing the details of the observed trends, the reader is reminded that laser pulses of different length employ different field amplitudes $\varepsilon^0_{t_p}$ (see Eq.~\ref{eq:pulse})  to carry the same total energy.

Within MCTDH, three different dynamical regimes are observed: i) the shortest pulse ($t_p$=2~fs) excites around 30\% of the population to higher-lying states,
ii)  laser pulses with a $t_p$ between 10 and 50~fs excite about 55\% of the population, and
iii) very long pulses beyond $t_p$ of 100~fs induce a diverging behavior, where the ground state is repopulated at later times where the laser is still active.
The reduced excitation in the dynamics including the very short $t_p=$2~fs laser pulse is due to the fact that the pulse carries only few cycles.
This results in a higher uncertainty for the energy and interference effects are increased as the amplitude of the laser field ($\varepsilon^0_{t_p}$) is adapted to pack the same amount of energy as the longer pulses in this short laser pulse.
When comparing the excitation process of the $t_p={10}$~fs laser pulse to the $t_p={17}$~fs, $t_p={30}$~fs or $t_p={50}$~fs dynamics, most of the effects hindering the excitation towards excited states present in the $t_p=2$~fs case are gone and an almost identical level of excited population is achieved.
This excitation of an identical amount of population is reminiscent of an ideal non-interacting case where dynamics in the excited states initiated at the start of the laser pulse does not enhance or hamper further excitation during the duration of the laser pulse.
When going to the $t_p={100}$~fs and $t_p={200}$~fs laser pulses, this ideal picture does not hold true anymore, and a more complex S$_0$ population behavior is observed.

\begin{figure}
\includegraphics{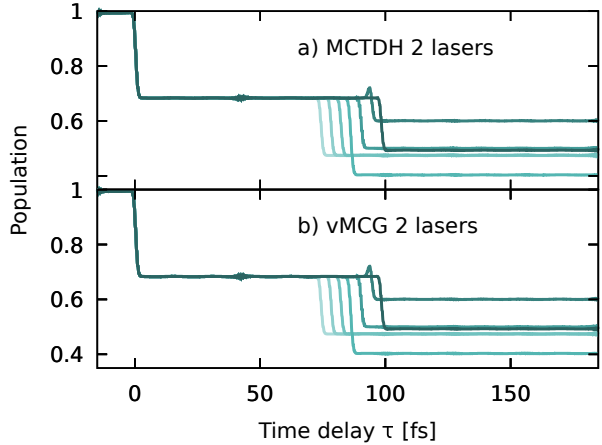}
\caption{\label{fig:so2_interception} Time-resolved S$_0$ population for (a) MCTDH and (b) vMCG with 50 GBFs dynamics in the presence of two short ($t_p=2$~fs) laser pulses, separated by a time-delay $\tau$. Every colored line represents the S$_0$ population of a different set of dynamics with increasing time-delay.}
\end{figure}

Inspection of the nuclear wave function for the dynamics using $t_p={100}$~fs and $t_p={200}$~fs lasers reveals that the diverging behavior is caused by the return parts of the excited state wave packet that re-enter the Franck-Condon region.
Once part of the wave packet is in the Franck-Condon region, additional interference terms arise as the returning excited wave packet and the remaining ground state wave function can interfere if the laser still couples the bright excited state with the ground state.
To verify whether this is actually the cause for the observed behavior, new sets of MCTDH simulations have been carried out where two identical $t_p=2$~fs lasers separated by a time interval $\tau$ are employed.
This way, one laser acts as a pump pulse and the other as probe, detecting whether the returning wave packet  causes interference terms.
The resulting populations for different $\tau$ delay times are shown in Fig.~\ref{fig:so2_interception}a.
It can be seen that the probe laser just excites more $S_0$ population for small $\tau$ values.
Upon reaching a time delay $\tau >$ 85~fs, a strong dependence on $\tau$ is found, which first enhances the excitation induced by the probe pulse but results in a reduced excitation for delays of 89, 93, 97~fs.
From this it can be concluded that after ca. 90~fs, a  recurrence of the excited wave packet takes place and further coupling with the laser field yields a new interference term, drastically altering the observed overall excitation process.

As the lasers up to $t_p={50}$~fs are too short in time to be still active when the wave packet returns, the excitation follows the observed pattern.
Only when going beyond this, a disturbance in the excitation pattern due to this effect is observed.
To further support this argument, a complex absorbing potential (CAP) has been employed in the MCTDH calculations that destroys the excited state wave packet once it leaves the Franck-Condon region after initial excitation, thus hindering a possible return.
Indeed, as seen in Fig.~\ref{fig:so2_s0}b, the S$_0$ populations do not show this interference term anymore.
Instead, the laser pulse induces a 55\% S$_0$ population inversion.

Figure~\ref{fig:so2_s0}c demonstrates that vMCG dynamics shows an extremely good agreement with MCTDH for laser pulses up to $t_p={50}$~fs.
Therefore, for short to medium laser pulses, vMCG is able to capture the most important parts of the excitation process induced by explicit laser pulses.
However, for longer laser pulses, vMCG deviates strongly from the MCTDH reference as the  behavior due to the interference with the returning wave packet observed for the S$_0$ population of the  $t_p={100}$~fs and $t_p={200}$~fs lasers are not reproduced.

Conducting the same pump-probe laser sequence dynamics as for MCTDH using two $t_p=2$~fs pulses resulted in a perfect agreement between the vMCG and the MCTDH dynamics (see Fig.~\ref{fig:so2_interception}b).
This shows that vMCG is well suited to describe the additional interference in the pump-probe setup, especially considering that only 50~GBFs were necessary to obtain the agreement. 
However, in simulations employing long laser pulses with $t_p={100}$~fs and $t_p={200}$~fs, vMCG does not capture the interference that was observed in MCTDH.
In these cases, the used GBFs are spread too thin, as the vMCG algorithm tries to describe the complete distributed excited state wave function at once --a much more difficult task than tracking the single excited state wave packet created in the pump-probe setup.
Exploratory calculations employing larger numbers of GBFs indicated that the reversal of the population flow occurring for the S$_0$ population of the $t_p={200}$~fs dynamics needs more than 125~GBFs to be even visible.
Unfortunately, calculations using this and larger number of GBFs could not be converged.

When comparing the MCTDH and vMCG dynamics to the FSSH dynamics using the $^{\mathrm{DIAG}}$EDC$_v^+$ set of parameters, see Fig.~\ref{fig:so2_s0}d, a general underestimation of about 5 to 10\% in the amount of excited population is observed for laser lengths up to $t_p={50}$~fs.
The reduced amount of excitation is mainly due to the employed decoherence correction.
Different from other decoherence corrections like AFSSH, the EDC acts instantaneously, meaning that in every time step of the simulations, the electronic coefficients of non-active states will be damped and the coefficient of the active state increased.
During the excitation process, the applied external field will periodically and continuously increase the electronic coefficient of excited states and decrease the coefficient in the ground state, thus enabling the chance for the trajectory to perform a surface hop.
The EDC therefore directly counteracts the influence of the laser field, resulting in an lower amount of excited population.
Interestingly, applying the same amount of energy over a longer time did not decrease the performance of the EDC correction further.
Similarly to vMCG, the S$_0$ populations in FSSH do not show any significant irregularities when moving to the $t_p={100}$~fs and the $t_p={200}$~fs lasers and instead correspond closely to the populations observed in MCTDH when employing a CAP.

\begin{figure*}
\includegraphics{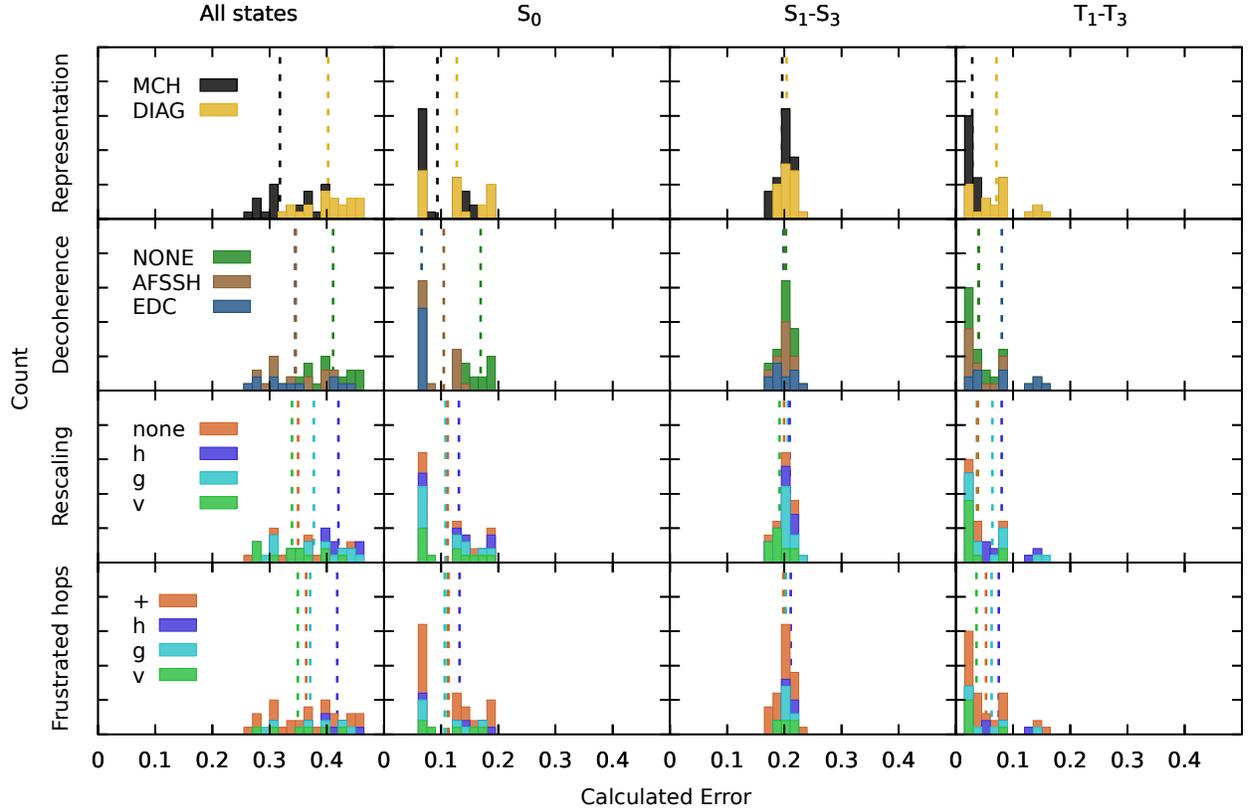}
\caption{\label{fig:so2_30fs_errors} Calculated $\epsilon$ values for each set of surface hopping parameters employed for the simulation of SO$_2$ represented as a stacked histogram, combining errors that fall into bins of size 0.015. The contribution of each surface hopping parameter towards the histogram is colored as indicated. Dashed lines represent the average error associated with the identical colored parameter for the given panel. Each row displays the contribution of all parameters associated with a specific setting, as specified. The left-hand side panel shows the combined error running over all considered states. The next panel shows $\epsilon_{S_0}$. The remaining columns show the $\epsilon^r_{sing}$ and $\epsilon^r_{trip}$ values. Note that all excited state populations have been renormalized to 1 in every time step for MCTDH and each FSSH simulation (see Eq.~\ref{eq:renorm_error}).}
\end{figure*}

A screening for all combinations of surface hopping parameters considered in Section~\ref{subsec:so2_nolaser} has been conducted using the $t_p={30}$~fs laser pulse, where no additional interference due to recurrence of the wave packet is observed.
Note that the total calculated error with respect to the reference MCTDH dynamics is now split into an error in the S$_0$ population, $\epsilon_{S_0}$, calculated according to Eq.~\ref{eq:error} while the error in the excited states, $\epsilon^r$, is calculated using renormalization following Eq.~\ref{eq:renorm_error}.
$\epsilon^r$ is again split up the contributions of singlets and triplet states. 
The total calculated error is then just the sum over these three contributions: $\epsilon=\epsilon_{S_0}+\epsilon^r_{sing}+\epsilon^r_{trip}$.
All calculated $\epsilon$ values for all states, the S$_0$, the excited singlet and the excited triplet states are collected in Fig.~\ref{fig:so2_30fs_errors} and Tables~SIV-SVII.
A very flat distribution of error values is observed for the total error, ranging from 0.265 to 0.462 showing an increase in overall error by 75\% for the worst set of parameters with respect to the best set.

First, the capability of the different parameters to describe the initial excitation is discussed.
When investigating the influence of the representation on $\epsilon_{S_0}$, the MCH picture is found to be better on average than the diagonal counterpart.

Regarding decoherence, the EDC performs best on average, giving very similar deviations from the MCTDH populations, irrespective of the chosen representation.
This consistently good performance is rather surprising following the observations above as the EDC in general leads to an underestimation of the excited population due to its instantaneous nature.
However, the ability to converge fast towards a trajectory that no longer contains considerable electronic S$_0$ population after the initial excitation is found to be beneficial here.
This lowered population in the S$_0$ of an already excited trajectory reduces the chance of stimulated emission and therefore increases the overall agreement with the MCTDH reference.
The use of AFSSH is found with a higher average error than the EDC.
All sets of parameters that use AFSSH paired with the MCH representation result in a $\epsilon_{S_0}$ between 0.058 to 0.076 and therefore show the lowest $\epsilon_{S_0}$ values.
However, when using AFSSH in combination with the diagonal representation, the errors essentially double to a range of 0.123 to 0.136, showing a strong sensitivity with the representation.
Such differences are not observed in the more robust EDC, indicating that the diagonal representation is interfering with the AFSSH algorithm in SO$_2$.

Using no decoherence correction at all is associated with a very bad agreement throughout all considered sets of parameters.
This indicates that the correct decay of the ground state electronic coefficient plays a crucial role once the trajectory is excited.
The performance in the absence of a treatment for the inherent overcoherence is slightly increased in the MCH representation and worsened in the diagonal one.

The different parameters for rescaling the kinetic energy after a hop and treating frustrated hops yielded almost identical averages.
Indeed, this behavior is expected as these options manipulate the excited state dynamics after the initial excitation has occurred and are therefore of limited importance to the excitation process itself.

We now analyze the dynamics in the excited states and see that a much more compact error distribution in error values is observed for the renormalized  $\epsilon^r_{sing}$ as compared to the dynamics in absence of a laser field. 
Almost all parameters give identical averages, with these averages also being smaller than for the dynamics without the laser field.
The lower average errors are mostly due to loss of the fine structure of the population curves in the $^1$B$_1$ and the $^1$A$_2$ states:
This fast and strongly oscillating behavior of the excited state populations depicted in Fig.~\ref{fig:so2_nolaser}a was one of the major sources of disagreement between MCTDH and FSSH, with FSSH predicting more damped oscillations.
Applying a long laser pulse  completely changes the observed picture, as the starting point of the dynamics in the excited states is now distributed across the duration of the complete laser pulse as can be seen in Fig.~S1 of the SI.
The delayed excitation results in a smearing of the transfer between the $^1$B$_1$ and the $^1$A$_2$ states that now show a simple oscillating behavior without fine structure.
This behavior is easier to reproduce using FSSH, which also profits from the very forgiving nature of using populations to calculate the deviation from exact quantum dynamics.
For the triplet states, all parameters that were found to increase description of triplet states in the simulations without explicit laser fields are also found to be the dominating factors in the presence of laser fields.
Hence, the MCH representation gives significantly improved results over the diagonal representation for the triplet state populations too.
The EDC is associated with higher errors in the triplet populations than using no decoherence correction or AFSSH.
For rescaling the energies after a hopping event and the treatment of frustrated hops, the velocity vector is the most prominent option followed by the complete neglect of adapting the kinetic energies.

Altogether, we find that an appropriate choice of the decoherence correction together with a compatible representation are the driving factors to capture the excitation process correctly. 
The EDC emerged here as a rather robust variant while AFSSH was found to perform better in the MCH basis.
For the remaining dynamics, the MCH basis was found to increase agreement for the triplet populations with the MCTDH reference.
Additionally, using the $\mathbf{h}$ and $\mathbf{g}$ vectors to rescale or reflect the velocity vector along after a real or frustrated hop was found to give larger deviations with respect to the reference.

\begin{figure}
\includegraphics{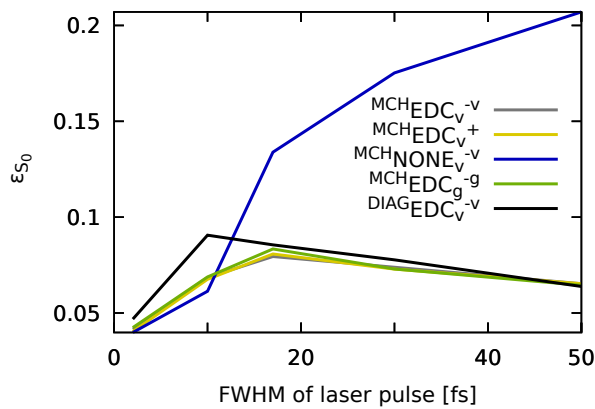}
\caption{\label{fig:so2_laserlength} Calculated error in the S$_0$ population for each set of FSSH parameters that resulted in the lowest average errors for at least one of the columns in Fig.~\ref{fig:so2_nolaser_errors} or Fig.~\ref{fig:so2_30fs_errors} for different $t_p$ of the laser pulse.}
\end{figure}

There are multiple approaches to find the best set of surface hopping parameters in the presented case: either the best performing set of parameters for a given $\epsilon$ is taken or the parameter of each option that obtained the lowest average error is considered.
In this work, the parameters yielding the lowest average errors have been taken for each option, as this represents a more robust error measure instead of cherry-picking a single combination that could well be the result of error compensation.
Therefore, the following sets have been determined in absence of a laser field: $^{\mathrm{MCH}}$NONE$_v^{-v}$ (best average $\epsilon$ and $\epsilon_{trip}$), $^{\mathrm{DIAG}}$EDC$_v^{-v}$ (best $\epsilon_{sing}$).
For dynamics initiated using a $t_p={30}$~fs laser, the combinations $^{\mathrm{MCH}}$EDC$_v^{-v}$ (best $\epsilon$), $^{\mathrm{MCH}}$EDC$_g^{-g}$ ($\epsilon_{S_0}$) and $^{\mathrm{MCH}}$EDC$_v^+$ ($\epsilon^r_{sing}$) were added, as the on average best performing options resulted again in  $^{\mathrm{MCH}}$NONE$_v^{-v}$ for $\epsilon^r_{trip}$.
These sets where then used to simulate dynamics using different laser lengths for excitation varying in size from $t_p=2$~fs to $t_p={50}$~fs and $\epsilon_{S_0}$ calculated with respect to the MCTDH dynamics.
The resulting deviations from the reference dynamics are depicted in Fig.~\ref{fig:so2_laserlength}.
Three different classes of surface hopping parameters can be clearly distinguished: (i) surface hopping sets using the MCH representation and the EDC clearly show the same trend for different $t_p$ of the laser, starting with a very small error that increases for $t_p={10}$~fs and $t_p={17}$~fs before decreasing for lasers $t_p={30}$~fs and  $t_p={50}$~fs long.
Essentially, no differences between the different rescaling or reflection parameters can be discerned for these three combinations.
(ii) Employing the diagonal representation initially results in a larger deviation from the MCTDH S$_0$ population for short laser lengths, which can be attributed to the different form the laser coupling takes in the diagonal representation.
For longer laser lengths this is evened out and no differences are observed between EDC in the diagonal and the MCH representation when looking at the S$_0$ populations.
(iii) A completely different picture can be observed when no decoherence correction is used ($^{\mathrm{MCH}}NONE_v^v$).
Here, the excitation due to very short laser pulses is well described but as soon as the excitation events and subsequent dynamics start occurring at the same time scale, the need to treat overcoherence is apparent.
Longer laser pulses lead to a larger deviation from the MCTDH S$_0$ population resulting in very larger errors for this set of parameters.
 
\subsection{2-Thiocytosine}

The substituted nucleobase 2-thiocytosine represents a much more challenging system for dynamics due to the increased number of normal modes with respect to SO$_2$.
Ab initio on-the-fly FSSH simulations and experimental results are available for 2-thiocytosine,\cite{mai_origin_2016} showing a significant triplet yield. 
Substituting the on-the-fly PES by the more rigid  LVC Hamiltonian was found previously to result in a reduced transfer towards the triplet states after 500~fs but to maintain all other essential features.\cite{plasser_highly_2019}
With this in mind, in order to test the effect of different FSSH settings, here we consider simulation times of up to 400~fs. 

\subsubsection{2-Thiocytosine dynamics in absence of a laser field}

\begin{figure}
\includegraphics{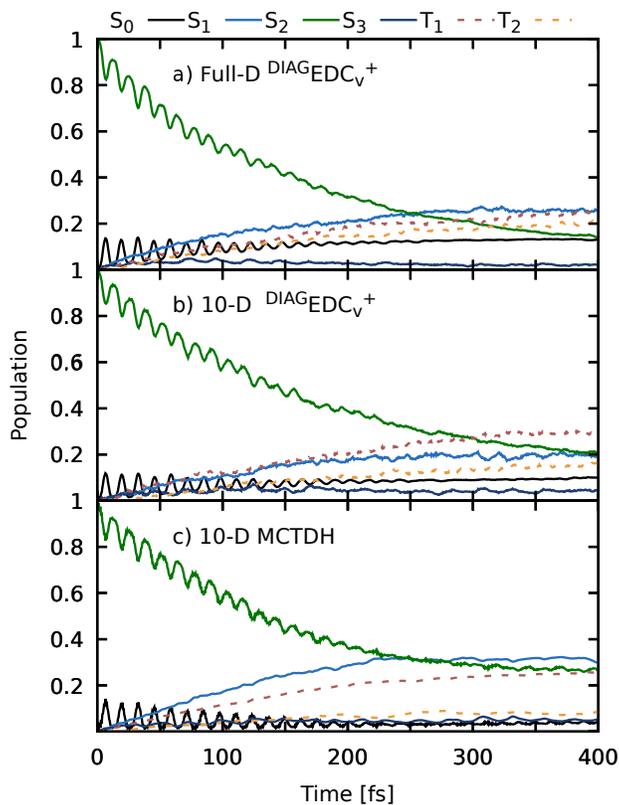}
\caption{\label{fig:2thio_modered} a) Diabatic populations of FSSH simulations on a full dimensional LVC model of 2-thiocytosine using the $^{\mathrm{DIAG}}$EDC$_v^v$ set of parameters starting from the bright S$_2$. Triplet populations (dashed) are summed up over all components of the respective triplet state. b) Diabatic populations obtained from $^{\mathrm{DIAG}}$EDC$_v^v$ simulations on a reduced 10-dimensional model of 2-thiocytosine starting again from the S$_2$ state. c) MCTDH dynamics on the 10-D model system, starting from the S$_2$ state.}
\end{figure}

Figure~\ref{fig:2thio_modered}a shows FSSH dynamics including all normal modes in conjunction with the $^{\mathrm{DIAG}}$EDC$_v^+$ set of parameters in the absence of a laser field.
The ground state Wigner distribution of geometries has been excited into the bright diabatic S$_2$ state from where subsequent dynamics takes place.
A strong coupling of this bright state with S$_0$ results in a fast oscillating transfer between these two states, recovering population in the S$_0$ state. 
At the same time, internal conversion to the diabatic S$_1$ and intersystem crossing to the triplet manifold is observed.
After 400 fs, both the S$_1$ and the S$_2$ have similar populations, and a higher population is found in the diabatic T$_1$ state.

In order to obtain a reference dynamics to assess different FSSH parameters using an affordable MCTDH computation, the amount of normal modes was reduced using the so-called SHARC-gym.~\cite{gomez_surface_2019}
In the SHARC-gym systematic calculations to identify the most important normal modes are carried out as follows.
For every vibrational normal mode, an independent FSSH simulation is conducted with all gradients and coupling elements associated with that mode set to zero.
The so-obtained populations are then compared to the reference populations in Fig~\ref{fig:2thio_modered}a by calculating $\epsilon$ values according to Eq.~\ref{eq:error}.
If the omission of a mode results in a large $\epsilon$, the mode carries essential coupling terms for the overall dynamics and should be included in the reduced set (see the $\epsilon$ values associated with the neglect of specific normal modes in Section~S2~A).
Using this approach, it is possible to find the modes relevant for the deactivation dynamics without any selection bias that could derive if the selection would be conducted based on the strength of specific coupling elements --a common approach in low-dimensional systems carrying limited amounts of states.~\cite{lasorne_automatic_2008,richings_can_2019}
Following the SHARC-gym we selected the ten most important modes that can reproduce the 33-dimensional dynamics of 2-thiocytosine.
This reduced 10-D set will be now used for the remainder of the simulations. 
As it can be seen in Fig.~\ref{fig:2thio_modered}b the FSSH dynamics using this 10-D Hamiltonian captures the overall behaviour of the full-D very reasonably. 

With this reduced model at hand, MCTDH calculations have been converged yielding the populations presented in Fig~\ref{fig:2thio_modered}c.
Compared to the $^{\mathrm{DIAG}}$EDC$_v^+$ presented set of parameters, the MCTDH simulations predict larger S$_1$ and smaller T$_2$ population.
The amount of S$_0$ population is lower in MCTDH, due to the inability of EDC to describe fast oscillations as it tries to end up in pure states.
Overall, the agreement between both methods is nevertheless quite reasonable, with an error $\epsilon$=0.214 for the $^{\mathrm{DIAG}}$EDC$_v^+$ set.

\begin{figure*}
\includegraphics{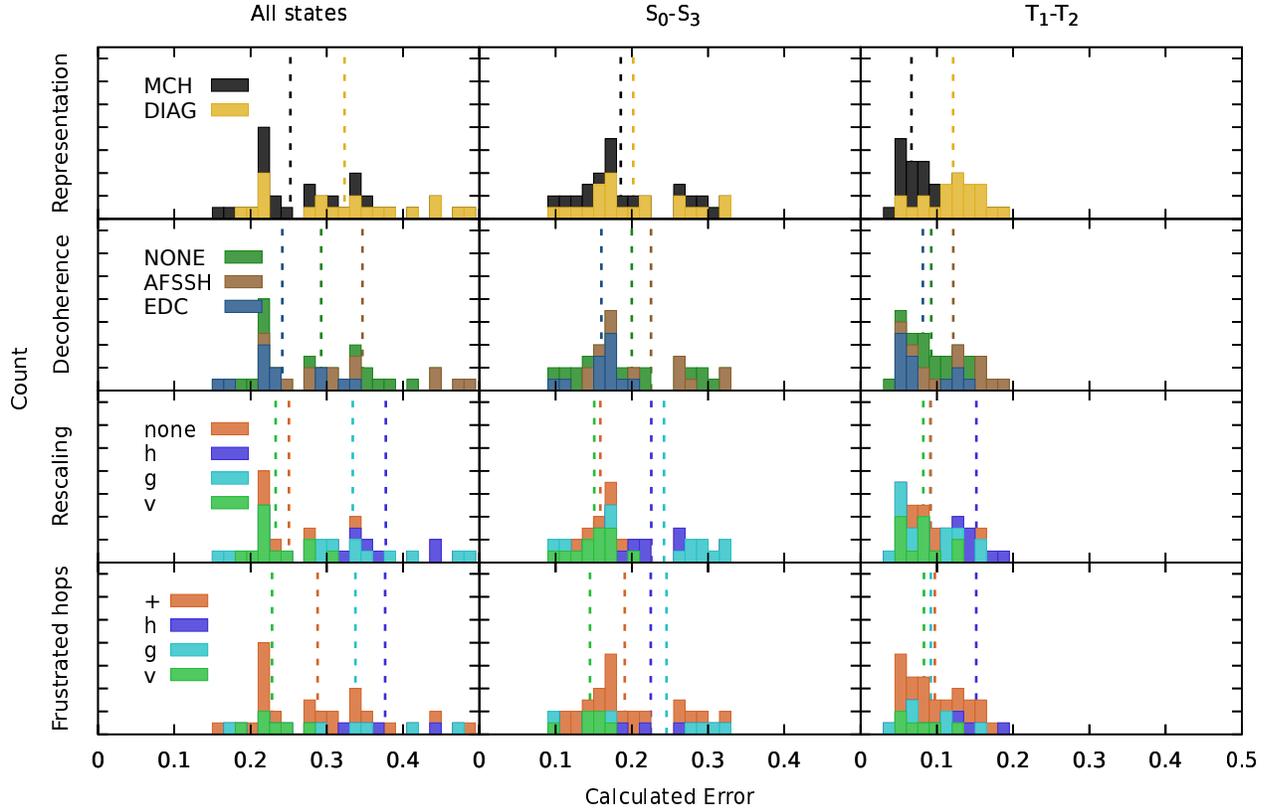}
\caption{\label{fig:2thio_nolaser_errors} The calculated error for the 10-D 2-thiocytosine dynamics for each set of surface hopping parameters is collected in this picture as a stacked histogram combining errors that fall into bins of size 0.015. Then the contribution of each surface hopping parameter towards the histogram is colored correspondingly. The dashed lines represent the average error associated with the identical colored parameter for the given panel. Each row displays the contribution of all parameters associated with a specific setting, labeled at the left hand side. The left-hand side panel shows the combined error running over all considered states. This error is then split up into the middle and the left-hand side panels, which display the calculated error over all singlet and triplet states respectively.}
\end{figure*}

Using the 10-D 2-thiocytosine model we obtain the distribution of $\epsilon$ values presented in Fig~\ref{fig:2thio_nolaser_errors} (see also Tables~SXII-XIV).
The errors are spread wide, ranging from 0.163 to 0.488, with the largest accumulation of parameter sets giving very good results.
Similar to SO$_2$, the MCH picture gives a better description of the triplet populations, while most simulations using the diagonal picture perform worse.
The EDC decoherence treatment performed well both for the singlet and the triplet populations. 
Using no decoherence correction resulted in somewhat surprisingly good results when compared to the AFSSH, which in turn gave larger $\epsilon$ values on average than the other treatments.
When including the effect of rescaling of the kinetic energies after a hop, the driving factor in the current dynamics becomes apparent: rescaling along the $\mathbf{h}$ and the $\mathbf{g}$ vectors is associated with the largest errors while neglecting the rescaling completely or rescaling along the velocity vector gives very good agreement with the MCTDH populations.
Interestingly, there are two exceptions to this, as the $^{MCH}EDC_g^{+}$ and the $^{MCH}EDC_g^g$ sets of parameters are the best performing sets investigated.
All but a single set of parameters using $\mathbf{v}$ gave better $\epsilon$ values than for any set using rescaling along $\mathbf{h}$.
Reflection of frustrated hops is found to be important; however, due to the pairing criterion of using the same vector for reflecting at a frustrated hop and to rescale along in the case of a hop, the errors are very similar to the errors observed for the corresponding rescaling option.
This does not hold true for not reflecting hops at all which was combined with all possible rescaling options.
The associated average error for not reflecting frustrated hops at all is found at almost the same value as the average over all calculated $\epsilon$ values , indicating that this parameter has no influence here.
The analysis of the sets resulting in larger $\epsilon$ values reveals that AFSSH dynamics in the diagonal representation results in too fast transfer from the S$_2$ to the $_1$ and increases population of the T$_1$ state, thus giving a bad description.
This effect is drastically enhanced when rescaling along  $\mathbf{h}$ or $\mathbf{g}$.

\subsubsection{2-thiocytosine dynamics in the presence of a laser field}

\begin{figure}
\includegraphics{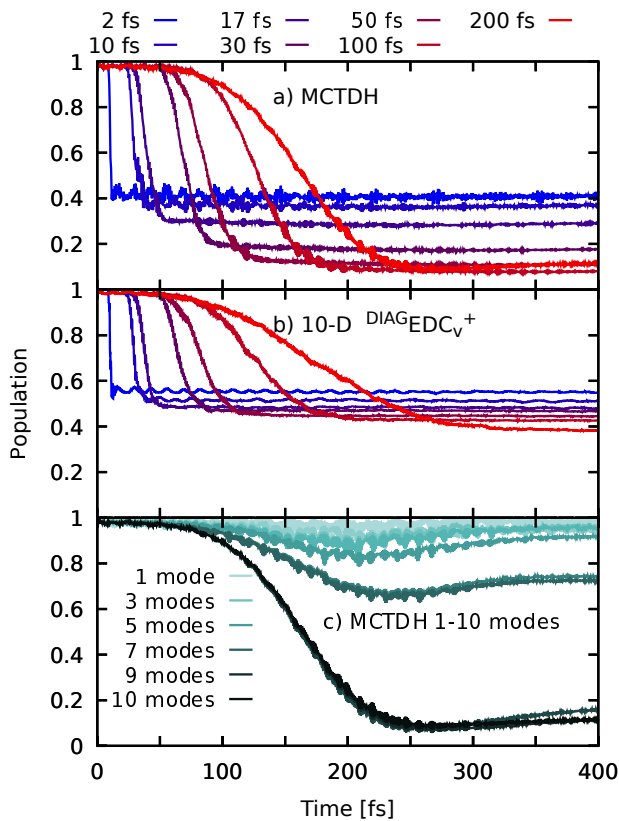}
\caption{\label{fig:2thio_s0} a)+b): Time-resolved S$_0$ population of 2-thiocytosine (10-D model) using 7 different laser pulse durations (as indicated)    
obtained with MCTDH (a) or FSSH with the parameter set $^{\mathrm{DIAG}}EDC_{v}^+$ (b). c) Time evolution of the S$_0$ population using the $t_p={200}$~fs and a different number of normal modes, from 1 to 10.}
\end{figure}

Following the framework presented in Section~\ref{subsec:so2_laser}, a set of 7 lasers that differ in their $t_p$ and their maximum amplitude have been applied to excite population from the ground state and initiate dynamics.
First, the influence of different lengths of laser excitation has been investigated.
Figure~\ref{fig:2thio_s0}a presents the results obtained for MCTDH.
Although the pulses are tuned to carry the same amount of overall energy, the amount of excited population increases with the pulse length.
Accordingly, the longest pulse achieves almost ground state population inversion. 
The same trend is achieved by the FSSH dynamics using the $^{\mathrm{DIAG}}$EDC$_v^+$ setup, see Fig~\ref{fig:2thio_s0}b,  but in a much weaker extent. 
This is in part due to the the EDC acting against the excitation process.
The hampering effect of the EDC is more pronounced than in SO$_2$ because the strength of this decoherence correction is based on the kinetic energy of the system, which on average is much larger for 2-thiocytosine than SO$_2$ due to size.

Intriguingly, the MCTDH S$_0$ populations do not show any surprises for longer pulses --contrary to what was observed for SO$_2$ where the recurrence of the wave packet resulted in additional interferences.
This is investigated in more detail in Fig.~\ref{fig:2thio_s0}c where ten different model systems containing one to ten normal modes of 2-thiocytosine (sorted according to their importance for the overall dynamics, see Table~SX for the explicit modes) are used to probe the interaction with the $t_p={200}$~fs laser pulse.
With very few modes, almost no population remains in the excited state as the recurrence of the wavepacket occurs very fast and further excitation is hampered and even reversed, similar to what occurred in SO$_2$.
The more modes are added to the system, the more excited state pathways open up, increasing the time needed until the excited wave packet returns to the Franck-Condon region, reducing the transfer back to the S$_0$ population as the laser duration is shorter than the recurrence time.
The increase in recoherence time shows that this effect is only affecting simulations of very small systems and it is almost undetectable for a 10-mode system with a rather long laser pulse of $t_p$ of 200~fs.

\begin{figure*}
\includegraphics{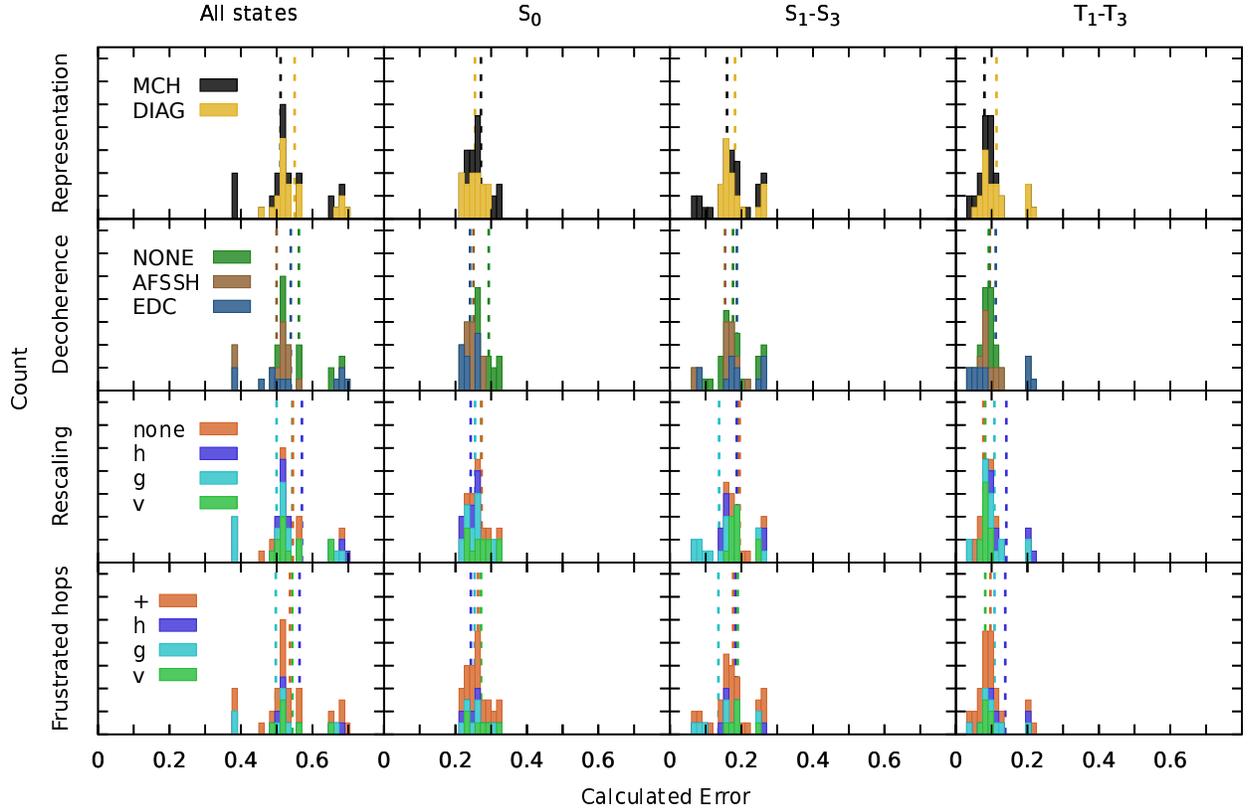}
\caption{\label{fig:2thio_30fs_errors} Calculated error for the 10-D 2-thiocytosine dynamics for each set of surface hopping parameters as a stacked histogram combining errors that fall into bins of size 0.015. The contribution of each surface hopping parameter towards the histogram is colored correspondingly. Dashed lines represent the average error associated with the identical colored parameter for the given panel. 
Each row displays the contribution of all parameters associated with a specific setting, labeled at the left hand side.
The left-hand side panel shows the combined error running over all considered states. The next panel shows $\epsilon_{S_0}$. The remaining columns show the $\epsilon^r_{sing}$ and $\epsilon^r_{trip}$ values. Note that all excited state populations have been renormalized to 1 in every time step for MCTDH and each FSSH simulation (see Eq.~\ref{eq:renorm_error}).}
\end{figure*}

The $t_p={30}$~fs pulse duration was selected to estimate the performance of the different FSSH parameter combinations on the 10 mode 2-thiocytosine model.
As can be seen in Fig.~\ref{fig:2thio_30fs_errors}, a very broad distribution of total errors is obtained for the overall error (see Tables~SXV-XVIII in the supporting information for the complete lists).
This is mostly due to the high deviation observed in the S$_0$ state population, yielding $\epsilon_{S_0}$ values larger than 0.2 for all sets.

The description of the S$_0$ population was found to be rather independent of the representation but strongly depending on the presence of a decoherence correction, with better results when this is considered. 
Rescaling the momenta after a surface hop impacts the S$_0$ population, with not rescaling the energies resulting in the bigger disagreement with MCTDH.
The overall increase in deviation from the MCTDH S$_0$ population when compared to the deviations in SO$_2$ can partially be attributed to the use of the rather strong laser pulse that inverts 90\% of the population in the MCTDH reference.
Indeed, strong laser fields have been  previously found to increase the deviation from quantum dynamics reference results.\cite{bajo_interplay_2014}
In that paper it was reported that the excitation process in general results in a net vibrational cooling of the swarm of excited trajectories with respect to the average kinetic energy of the ground state population, resulting in slower movement of the FSSH trajectories away from the region where the laser couples the bright and the ground state.
In the presence of a strong laser field, dwelling in the region of strong coupling increases the chance of inducing radiative emission to the ground state, effectively decreasing the amount of excited population.
This poses an important dilemma for carrying out FSSH dynamics with laser pulses when generating initial conditions: On the one hand side, the stronger the pulse, the more trajectories will be excited, thereby decreasing the amount of unexcited trajectories that are now obsolete for describing excited state dynamics and thus represent computation time essentially wasted.
On the other hand, stronger lasers will lead to larger deviations from the exact dynamics due to the  cooling effect and the increased interplay between different parts of the wave packet that are more strongly coupled.

For the excited state dynamics following excitation, $\epsilon^r_{sing}$ shows a rather narrow distribution.
Surprisingly, a complete change in the observed order of importance for the various parameters can be detected with respect to the dynamics without any external field.
Two effects are responsible for the presented errors: 
First, the influence due to the fast oscillating coupling element with the S$_0$ on the $\epsilon$ values is reduced as the S$_0$ component of this error is included in $\epsilon_{S_0}$ and therefore only taken into account via the S$_2$ state. 
Second, the fine structure is smoothed out, giving rise to simpler decay patterns as can be seen in Fig.~S~3 in the supporting information.
Still, it comes as a big surprise to find that the performance of the EDC is reduced and that rescaling of the kinetic energy after a hop along $\mathbf{g}$ in the MCH representation gives by far the best renormalized singlet populations.
The latter behaviour is the opposite of what was observed in absence of a laser field, i.e. this rescaling option was found with the largest average $\epsilon_{sing}$.
When including an explicit laser pulse, rescaling along $\mathbf{g}$ gives good agreement with MCTDH independent of the applied decoherence correction if the MCH representation is used.
However, a strong dependence on the applied decoherence correction in the diagonal representation can be seen where $^{\mathrm{DIAG}}EDC_g^{-g}$ and $^{\mathrm{DIAG}}EDC_g^+$ are among the largest deviations from the MCTDH results for all sets observed.

The deviation in the populations of the triplet states follows the same trends found in the excited state dynamics without an external field, with a decrease in the importance of the chosen representation and an overall better performance of the AFSSH. 
Rescaling along the nonadiabatic coupling vector is still found to decrease the agreement with the MCTDH reference.

Summarizing this section, modelling the excitation of S$_0$ population by an explicit laser pulse with $t_p=30$~fs in the 10-D 2-thiocytosine model system using FSSH results in deviations of about 30\%. 
Differences due to the representation are negligible while using a decoherence correction was found to be an important factor, similar to observations in SO$_2$.
For the excited state dynamics following excitation, rescaling of the energies after a hop was found to be important,
this time favoring rescaling along the gradient difference vector for the dynamics in the singlet manifold.
Transfer of population towards the triplet states is found to be only slightly influenced by the decoherence treatment and the chosen representation.


\section{Conclusions}
In this work, we benchmark the performance of different parameters that are needed in FSSH against a MCTDH reference for two molecular systems: SO$_2$ and 2-thiocytosine using parametrized LVC potentials.
We  investigate the effect of such parameters on simulations that do not explicitly include a laser field and compare with dynamics initiated by laser pulses of different time duration. 

Previous work\cite{zhou_nonadiabatic_2020,fiedlschuster_floquet_2016,fiedlschuster_surface_2017,bajo_interplay_2014,mignolet_excited-state_2019} has reported the inability of FSSH to correctly describe the excitation process and the subsequent dynamics in the presence of explicit light-matter interactions, as well as pointed out the dependence of the results on the chosen FSSH parameters. 
In this work, we quantitatively measure the deviation of FSSH from real wave packet dynamics, and by studying the system with and without the influence of external fields, we discern between effects coming from the dynamics itself or from the laser interaction.
Furthermore, the two chosen models SO$_2$ and 2-thiocytosine, with 3 and up 10 normal modes, are beyond the commonly  1-dimensional test systems employed to investigate the interaction with an external field.
The consideration of both singlet and triplet states pose an additional difficulty for FSSH as both very weak, delocalized coupling between singlets an triplets as well as the time-dependent coupling via an external field have to be treated correctly. 

We find that FSSH has difficulties in low-dimensional models if long laser pulses (FWHM>100~fs) are present, where the excited state wave packet returns to the Franck-Condon region within the time scale of the applied laser pulse. 
MCTDH simulations in SO$_2$ indicate that the returning wave packet interacts with the active laser pulse, resulting in an additional interference, for laser pulses of at least $t_p=100$~fs.
FSSH is not capable to reproduce these interference terms and vMCG can ony capture this effect when going beyond 100~GBFs.
Moving to larger model systems decreases the recurrence time until it is not observed with the longest  ($t_p$=200~fs) laser pulse in the 10-mode model of 2-thiocytosine, indicating that these effects will be absent in systems with more degrees of freedom.

In SO$_2$, no perfect set of parameters was found suitable to describe quantitatively all different transition processes triggered by an ultrafast ($t_p$=30~fs) laser pulse or when the simulation starts directly from the bright state. 
The decoherence correction emerges as the most essential parameter to model 
the excitation process and therefore also the overall laser-induced dynamics in the presence of 
laser pulses longer than a few cycles. 
The MCH representation where both the coupling via a laser field as well as the spin-orbit coupling are treated as off-diagonal elements, is found beneficial to describe adequately both the S$_0$ and the triplet populations.
This is mainly due to a better performance of AFSSH in the MCH representation to describe the change in the S$_0$ population as compared to AFSSH in the fully diagonal representation.
For the EDC, no such strong basis dependence was identified for the excitation process, but larger disagreements for the triplet states are observed.
For the observed dynamics in the excited singlet states, most parameters where found with almost identical errors.
When using a reasonable combination of FSSH parameters, similar deviations are obtained for the excitation process for different lengths of the laser pulse until the wave packet recurrence is observed.
Finally, the importance of rescaling the kinetic energy after a hopping event (not influenced by the laser field) is evident in the excited state dynamics, with rescaling along both the non-adiabatic coupling vector and the gradient difference vector showing high deviations for the triplet state dynamics.
Overall, the best FSSH parameter sets are associated with ca 20\% deviation in the amount of excited S$_0$ population with respect to MCTDH, while the subsequent population evolution is described with a similar level of accuracy as without the laser pulse.

For 2-thiocytosine, larger differences in the amount of excited population are observed due to the strong laser applied, increasing the difference in excited population by 35\% and more, almost irrespective of the chosen parameters.
The use of a decoherence correction  increases the overall agreement throughout the excitation process, similar to the observations in SO$_2$.
In the subsequent dynamics, rescaling of the kinetic energy after a hop reveals itself as the most important factor to obtain lower errors with respect to the MCTDH populations.
Interestingly, the gradient difference vector is found to mostly result in large deviations from the reference if no laser is present but becomes the best performing parameter in the dynamics including the $t_p=30$~fs laser pulse.

In conclusion, this work shows the difficulties in choosing a universal set of parameters that guarantees quantitative agreement against quantum reference results, particularly in the presence of an electric field.
However, and despite many difficulties, FSSH can qualitatively reproduce the dynamics ensuing after excitation using an exemplary laser pulse of FWHM=30~fs on two different test systems, with three and ten dimensions --an encouraging result to pave the way to use of explicit laser pulses in FSSH simulations.
Based on the observations made in this paper, a few caveats have been identified. 
The use of very long laser pulses is discouraged due to the increased chance of quantum interferences occurring, especially for small systems.
The same holds true for the use of very strong laser pulses, which will enhance problems inherent to FSSH without any laser pulse.
Additionally, it is advised to use a decoherence correction (from the many available) because it is demonstrated to be in most cases the dominating factor to increase the agreement with the MCTDH reference, especially when considering explicit excitation.
Caution is also advised when it comes to choosing a vector to adjust the momenta after a non-laser induced hop, as this parameter strongly influences the presented dynamics.

\section{Supporting Information}

Examples of population evolution in the presence of laser fields, highlighting the renormalization conducted within Eq.~\ref{eq:renorm_error}. Sorted lists containing all calculated $\epsilon$ values used in FIG.s~\ref{fig:so2_nolaser_errors},\ref{fig:so2_30fs_errors},\ref{fig:2thio_nolaser_errors}, and \ref{fig:2thio_30fs_errors}. All $\epsilon$ values calculated for SO$_2$ using vMCG with different numbers of basis functions. (PDF)

Input and operator files used to simulate all MCTDH and vMCG calculations in QUANTICS. Molden files and LVC template files for SO$_2$, 2-thiocytosine (full-D and 1- up to 10-D) use in SHARC. (ZIP)

\bibliography{main.bib}

\end{document}


\preprint{AIP/123-QED}

\title[Supporting Information: Validating Fewest-Switches Surface Hopping in the Presence of Laser Fields]{}

\author{Moritz Heindl}
\author{Leticia Gonz\'alez}
 \email{leticia.gonzalez@univie.ac.at}
\affiliation{Institute of Theoretical Chemistry, Faculty of Chemistry, University of Vienna, Währingerstr. 17, 1090 Vienna, Austria
}
\begin{center}
    
{\huge \bf Supporting Information: \\
Validating Fewest-Switches Surface Hopping in the Presence of Laser Fields}
\end{center}

Moritz Heindl$^1$ and Leticia González$^1$

\textit{$^1$Institute of Theoretical Chemistry, Faculty of Chemistry, University of Vienna, Währingerstr. 17, 1090 Vienna, Austria}

\date{\today}

\section{SO$_2$}

\subsection{Population dynamics with laser fields} \label{sec:so2_pop}
\begin{figure}
\includegraphics{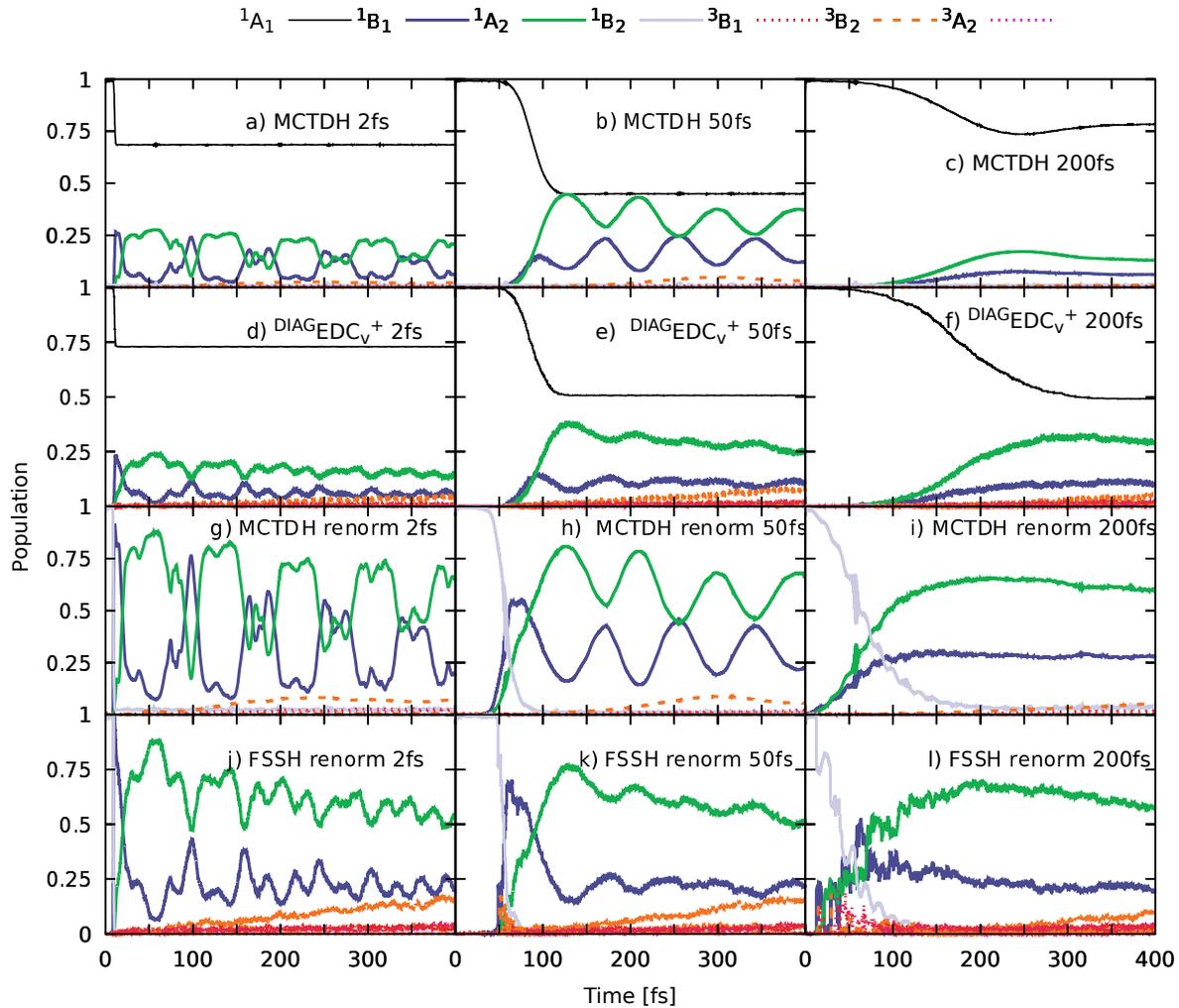}
\caption{\label{fig:so2_laserpops} Evolution of the populations in all considered states in the SO$_2$ molecule in presence of laser fields. a-c) MCTDH populations for dynamics using varying length ($t_P$) of the laser pulse. d-f) FSSH simulations using the $^{\mathrm{DIAG}}$EDC$_\mathrm{v}^+$ set of parameters for the same laser lengths already employed in a-c). g-i) Renormalized excited state populations for the MCTDH dynamics shown in a-c) where each excited state population is adapted according to $p^r_i=\frac{p_i}{1-p_1}$ where $p_1$ denotes the ground state population and $p_i$ the population in the $i$th excited state in a-c). j-l) Renormalized excited state populations for the FSSH dynamics shown in d-f).}
\end{figure}

The change in the overall population dynamics due to the excitation using a laser pulse with finite length is shown in FIG~\ref{fig:so2_laserpops} where the populations for dynamics in presence of laser fields of length $t_p$ 2~fs, 50~fs, and 200~fs are shown for both MCTDH and a specific FSSH setup. 
The loss of the fine structure of the population dynamics is directly visible with the $t_p=2$~fs dynamics showing a lot of population changes with distinctly peaked populations, this behavior is reduced down to a simple oscillatory movement for $t_p=50$~fs. 
For the longest applied laser pulse, no fine structure at all is visible as the populations evolve smoothly.
FIG~\ref{fig:so2_laserpops}g-l) show how the population curves shown in FIG~\ref{fig:so2_laserpops} look like after renormalization according to the two fractions in Equation~10.
The final renormalized error $\epsilon^r$ is obtained by calculating the differences in population between e.g. FIG~\ref{fig:so2_laserpops}g and FIG~\ref{fig:so2_laserpops}j.
As can be seen here, this renormalization process allows to directly compare the excited state populations but suffers from noise in the beginning of the evaluation of the error, where almost all population is still in the S$_0$ state.
This can be best understood by looking at FIG~\ref{fig:so2_laserpops}i where initially all renormalized population is in the $^1$B$_2$ state.
The $^1$B$_2$ state is only marginally populated in the real (=not renormalized) populations, as only a coupling element between the ground state with this highest singlet state results in a population of about 0.004.
The small value for $p_i$ is ballooned to a very large value of about 1 in the corresponding renormalized population ($p_i^r$) in the beginning of the dynamics as no other excited state is populated.
Therefore, very small deviations in the population of the $^1$B$_2$ state in the beginning of the dynamics result in large $\epsilon$ values.
To minimize the influence of this noise, an arbitrary cut-off value was introduced and contributions to $\epsilon^r$ were only calculated if the S$_0$ population was below 0.98.

\clearpage

\subsection{$\epsilon$ values}

\begin{table}[b]
\scriptsize
\begin{tabular}{ccccc}
\hline
& $\epsilon$ & $\epsilon_{sing}$ & $\epsilon_{trip}$ & \\
\hline
$^\mathrm{DIAG}$NONE$_{v}^{+}$  & 0.2402        & 0.2097        & 0.0306        & \\
$^\mathrm{MCH}$AFSSH$_{v}^{+}$  & 0.2422        & 0.2117        & 0.0305        & \\
$^\mathrm{MCH}$EDC$_{none}^{+}$ & 0.2433        & 0.2138        & 0.0296        & \\
$^\mathrm{MCH}$EDC$_{v}^{+}$    & 0.2448        & 0.2123        & 0.0324        & \\
$^\mathrm{DIAG}$EDC$_{v}^{+}$   & 0.2478        & 0.2081        & 0.0398        & \\
$^\mathrm{MCH}$NONE$_{v}^{+}$   & 0.2483        & 0.2130        & 0.0352        & \\
$^\mathrm{MCH}$EDC$_{v}^{v}$    & 0.2535        & 0.2214        & 0.0321        & \\
\hline
\end{tabular}
\caption{$\epsilon$ values obtained for all sets of considered FSSH dynamics in SO$_2$ without any external laser field using MCTDH as reference. The methods are sorted according to lowest overall $\epsilon$.}
\label{tab:so2_epsilons_tot}
\end{table}

\begin{table}
\scriptsize
\begin{tabular}{ccccc}
\hline
& $\epsilon$ & $\epsilon_{sing}$ & $\epsilon_{trip}$ & \\
\hline
$^\mathrm{DIAG}$EDC$_{v}^{+}$   & 0.2478        & 0.2081        & 0.0398        & \\
$^\mathrm{DIAG}$NONE$_{v}^{+}$  & 0.2402        & 0.2097        & 0.0306        & \\
$^\mathrm{MCH}$AFSSH$_{v}^{+}$  & 0.2422        & 0.2117        & 0.0305        & \\
$^\mathrm{MCH}$EDC$_{v}^{+}$    & 0.2448        & 0.2123        & 0.0324        & \\
$^\mathrm{MCH}$NONE$_{v}^{+}$   & 0.2483        & 0.2130        & 0.0352        & \\
$^\mathrm{MCH}$EDC$_{none}^{+}$ & 0.2433        & 0.2138        & 0.0296        & \\
$^\mathrm{DIAG}$AFSSH$_{v}^{+}$ & 0.2561        & 0.2142        & 0.0419        & \\
\hline
\end{tabular}
\caption{Identical information to TABLE~\ref{tab:so2_epsilons_tot}. The methods are sorted according to lowest $\epsilon_{sing}$.}
\label{tab:so2_epsilons_sing}
\end{table}

\begin{table}
\scriptsize
\begin{tabular}{ccccc}
\hline
& $\epsilon$ & $\epsilon_{sing}$ & $\epsilon_{trip}$ & \\
\hline
$^\mathrm{DIAG}$NONE$_{v}^{v}$  & 0.2704        & 0.2410        & 0.0295        & \\
$^\mathrm{MCH}$EDC$_{none}^{+}$ & 0.2433        & 0.2138        & 0.0296        & \\
$^\mathrm{MCH}$EDC$_g^{+}$      & 0.3003        & 0.2702        & 0.0301        & \\
$^\mathrm{MCH}$AFSSH$_{v}^{+}$  & 0.2422        & 0.2117        & 0.0305        & \\
$^\mathrm{DIAG}$NONE$_{v}^{+}$  & 0.2402        & 0.2097        & 0.0306        & \\
$^\mathrm{MCH}$EDC$_{v}^{v}$    & 0.2535        & 0.2214        & 0.0321        & \\
$^\mathrm{MCH}$EDC$_{v}^{+}$    & 0.2448        & 0.2123        & 0.0324        & \\
\hline
\end{tabular}
\caption{Identical information to TABLE~\ref{tab:so2_epsilons_tot}. The methods are sorted according to lowest $\epsilon_{trip}$.}
\label{tab:so2_epsilons_trip}
\end{table}

\begin{table}
\scriptsize
\begin{tabular}{cccccc}
\hline
& $\epsilon$ & $\epsilon_{S_0}$ & $\epsilon^r_{sing}$ & $\epsilon^r_{trip}$ & \\
\hline
$^\mathrm{MCH}$EDC$_{none}^{+}$ & 0.2653        & 0.0604        & 0.1761        & 0.0288        & \\
$^\mathrm{MCH}$AFSSH$_{v}^{+}$  & 0.2714        & 0.0714        & 0.1743        & 0.0256        & \\
$^\mathrm{MCH}$EDC$_{v}^{+}$    & 0.2728        & 0.0640        & 0.1783        & 0.0305        & \\
$^\mathrm{MCH}$EDC$_{v}^{v}$    & 0.2737        & 0.0648        & 0.1802        & 0.0286        & \\
$^\mathrm{MCH}$AFSSH$_g^g$      & 0.2892        & 0.0643        & 0.1990        & 0.0260        & \\
$^\mathrm{MCH}$EDC$_g^g$        & 0.3002        & 0.0641        & 0.2021        & 0.0341        & \\
$^\mathrm{MCH}$AFSSH$_g^{+}$    & 0.3011        & 0.0638        & 0.2145        & 0.0228        & \\
\hline
\end{tabular}
\caption{$\epsilon$ values obtained for all sets of considered FSSH dynamics in SO$_2$ including a $t_p=30$~fs laser field using MCTDH as reference. The methods are sorted according to lowest overall $\epsilon$}
\label{tab:so2_epsilons_laser_tot}
\end{table}

\begin{table}
\scriptsize
\begin{tabular}{cccccc}
\hline
& $\epsilon$ & $\epsilon_{S_0}$ & $\epsilon^r_{sing}$ & $\epsilon^r_{trip}$ & \\
\hline
$^\mathrm{MCH}$EDC$_{none}^{+}$ & 0.2653        & 0.0604        & 0.1761        & 0.0288        & \\
$^\mathrm{MCH}$EDC$_g^{+}$      & 0.3114        & 0.0633        & 0.2093        & 0.0389        & \\
$^\mathrm{MCH}$AFSSH$_g^{+}$    & 0.3011        & 0.0638        & 0.2145        & 0.0228        & \\
$^\mathrm{MCH}$EDC$_{v}^{+}$    & 0.2728        & 0.0640        & 0.1783        & 0.0305        & \\
$^\mathrm{DIAG}$EDC$_{none}^{+}$        & 0.3288        & 0.0640        & 0.1847        & 0.0801        & \\
$^\mathrm{MCH}$EDC$_g^g$        & 0.3002        & 0.0641        & 0.2021        & 0.0341        & \\
$^\mathrm{MCH}$AFSSH$_g^g$      & 0.2892        & 0.0643        & 0.1990        & 0.0260        & \\
\hline
\end{tabular}
\caption{Identical information to TABLE~\ref{tab:so2_epsilons_laser_tot}. The methods are sorted according to lowest $\epsilon_{S_0}$.}
\label{tab:so2_epsilons_laser_s0}
\end{table}

\begin{table}
\scriptsize
\begin{tabular}{cccccc}
\hline
& $\epsilon$ & $\epsilon_{S_0}$ & $\epsilon^r_{sing}$ & $\epsilon^r_{trip}$ & \\
\hline
$^\mathrm{MCH}$AFSSH$_{v}^{+}$  & 0.2714        & 0.0714        & 0.1743        & 0.0256        & \\
$^\mathrm{MCH}$EDC$_{none}^{+}$ & 0.2653        & 0.0604        & 0.1761        & 0.0288        & \\
$^\mathrm{MCH}$NONE$_{v}^{+}$   & 0.3513        & 0.1460        & 0.1781        & 0.0272        & \\
$^\mathrm{MCH}$EDC$_{v}^{+}$    & 0.2728        & 0.0640        & 0.1783        & 0.0305        & \\
$^\mathrm{MCH}$EDC$_{v}^{v}$    & 0.2737        & 0.0648        & 0.1802        & 0.0286        & \\
$^\mathrm{DIAG}$NONE$_{v}^{+}$  & 0.3940        & 0.1814        & 0.1840        & 0.0286        & \\
$^\mathrm{DIAG}$EDC$_{v}^{+}$   & 0.3407        & 0.0683        & 0.1844        & 0.0880        & \\
\hline
\end{tabular}
\caption{Identical information to TABLE~\ref{tab:so2_epsilons_laser_tot}. The methods are sorted according to lowest $\epsilon^r_{sing}$.}
\label{tab:so2_epsilons_laser_sing}
\end{table}

\begin{table}
\scriptsize
\begin{tabular}{cccccc}
\hline
& $\epsilon$ & $\epsilon_{S_0}$ & $\epsilon^r_{sing}$ & $\epsilon^r_{trip}$ & \\
\hline
$^\mathrm{MCH}$NONE$_g^{+}$     & 0.3697        & 0.1411        & 0.2061        & 0.0225        & \\
$^\mathrm{MCH}$AFSSH$_{v}^{v}$  & 0.3060        & 0.0758        & 0.2075        & 0.0227        & \\
$^\mathrm{MCH}$AFSSH$_g^{+}$    & 0.3011        & 0.0638        & 0.2145        & 0.0228        & \\
$^\mathrm{DIAG}$AFSSH$_{none}^{+}$      & 0.3619        & 0.1247        & 0.2128        & 0.0244        & \\
$^\mathrm{DIAG}$NONE$_{v}^{v}$  & 0.4202        & 0.1776        & 0.2179        & 0.0247        & \\
$^\mathrm{DIAG}$AFSSH$_{v}^{v}$ & 0.3600        & 0.1285        & 0.2066        & 0.0249        & \\
$^\mathrm{MCH}$AFSSH$_{v}^{+}$  & 0.2714        & 0.0714        & 0.1743        & 0.0256        & \\
\hline
\end{tabular}
\caption{Identical information to TABLE~\ref{tab:so2_epsilons_laser_tot}. The methods are sorted according to lowest $\epsilon^r_{trip}$.}
\label{tab:so2_epsilons_laser_trip}
\end{table}
\clearpage

\subsection{vMCG deviation}

\begin{table}[h]
\begin{tabular}{cccc}
\hline
GBFs & $\epsilon$ & $\epsilon_{sing}$ & $\epsilon_{trip}$\\
\hline
10 & 0.2041 & 0.1754 & 0.0287   \\
25 & 0.1428 & 0.1049 & 0.0380   \\
51 & 0.0944 & 0.0603 & 0.0341  \\
75 & 0.0693 & 0.0367 & 0.0326  \\
100 & 0.0246 & 0.0164 & 0.0082 \\

\hline
\end{tabular}
\caption{$\epsilon$ values obtained for all vMCG dynamics in SO$_2$, employing different numbers of GBFs  without any external laser field using MCTDH as reference. The methods are sorted according to lowest overall $\epsilon$. The calculation using 50~GBFs did not conserve the norm of the wave function and was therefore replaced by a simulation using 51~GBFs.}
\label{tab:so2_vmcg_nolas}
\end{table}

\begin{table}[h]
\begin{tabular}{ccccc}
\hline
GBFs & $\epsilon$ & $\epsilon_{S_0}$ & $\epsilon^r_{sing}$ & $\epsilon^r_{trip}$\\
\hline
10& 0.3457 & 0.0176 & 0.2538  & 0.0744 \\
25& 0.2194 & 0.0464 & 0.1455  & 0.0275 \\
50& 0.1647 & 0.0108 & 0.1296  & 0.0242 \\
75& 0.1116 & 0.0045 & 0.0782  & 0.0289 \\
100& 0.0808 & 0.0016 & 0.0529 & 0.0263 \\
\hline
\end{tabular}
\caption{$\epsilon$ values obtained for all vMCG dynamics in SO$_2$, employing different numbers of GBFs  including a $t_p=30$~fs laser field and  using MCTDH as reference. The methods are sorted according to lowest overall $\epsilon$.}
\label{tab:so2_vmcg_nolas}
\end{table}
\newpage

\section{2-thiocytosine}

\subsection{Mode reduction}

Running MCTDH directly on the full-dimensional (33 vibrational degrees of freedom) 2-thiocytosine model is not viable/possible due to the increase in computational cost with higher number of degrees of freedom.
Hence, in this work, the full-D model is shrunk to a size, that can be successfully treated using MCTDH, using the SHARC-gym methodology described in Ref~\citenum{gomez_surface_2019}.
An alternative solution would be the use of multi-layer MCTDH which has been employed successfully in the investigation of even larger molecules\cite{wang_multilayer_2003,vendrell_multilayer_2011}.

\begin{figure}[h]
\includegraphics{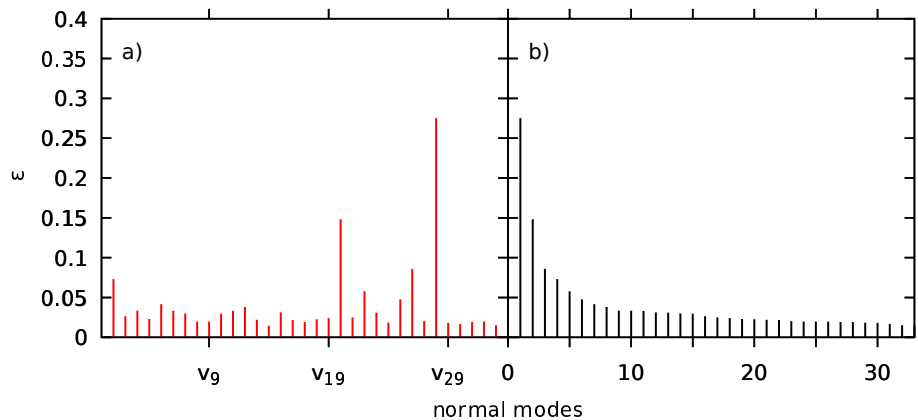}
\caption{\label{fig:2thio_modes} a) $\epsilon$ values for FSSH dynamics using the $^{\mathrm{DIAG}}$EDC$_\mathrm{v}^+$ set of parameters on a template of 2-thiocytosine where a single mode has been deactivated with respect to the full dimensional population evolution. The corresponding $\epsilon$ value is plotted against the number of the deactivated mode. b) same information as in a) but now sorted according to the corresponding $\epsilon$ value. The x-label now represents a sorting index.}
\end{figure}

In the SHARC-gym approach, FSSH dynamics are used to determine vibrational normal modes that are essential to describe the excited state dynamics, which can then be used to form a lower-dimensional Hamiltonian still capable of describing the overall excited state dynamics.
For this, a reference FSSH simulation using the full-D LVC model employing the $^{\mathrm{DIAG}}$EDC$_\mathrm{v}^+$ set of parameters is created, where the initially excited state was chosen to be the second excited state at the reference geometry.
All initial electronic populations and initial states of each of the 1000 trajectories was adapted to start in the corresponding state for each initial geometry.
Subsequently, 33 new simulations, each employing 1000 new trajectories starting in the same electronic state but each neglecting a different single vibrational mode (all couplings and gradients associated with this mode in the LVC template were set to zero) have been conducted.
For each of these dynamics, $\epsilon$ values according to Equation~9 in the main manuscript with respect to the full-D reference simulation have been calculated as listed in TABLE~\ref{tab:2thio_singlemodes} and can be seen in FIG~\ref{fig:2thio_modes}.
As can be seen there, most modes are associated with $\epsilon$ values below 0.05 and therefore result in almost identical dynamics.
After sorting the modes based on the associated error that is introduced when this mode is omitted, an arbitrary cut-off of 10 modes has been chosen that will be included in the reduced LVC model.
When comparing the 10-D FSSH dynamics to the full-D dynamics, an $\epsilon$ of 0.219 is obtained (for the populations see FIG~7 in the main manuscript). For the list of modes used in all of the 1- to 10-D model systems and the associated $\epsilon$ values, see TABLE~\ref{tab:2thio_lowerd}.
As can be seen there, the $epsilon$ values for the 1- to 10-D do not converge smoothly but instead show large gaps for example at the 4-D model.
These large jumps are caused when specific modes open up the avenue to populate specific states, from where some previously included modes can transfer population even further to other states.
These transfers where hampered previously as the state, from where population is flowing from was not populated before.

\begin{table}[b]
\scriptsize
\begin{tabular}{cc}
\hline
Vibrational mode& $\epsilon$ \\
\hline
$v_{28}$ & 0.275  \\
$v_{20}$ & 0.148  \\
$v_{26}$ & 0.086  \\
$v_{1}$ & 0.073  \\
$v_{22}$ & 0.058  \\
$v_{25}$ & 0.048  \\
$v_{5}$ & 0.042  \\
$v_{12}$ & 0.038  \\
$v_{6}$ & 0.033  \\
$v_{3}$ & 0.033  \\
$v_{11}$ & 0.033  \\
$v_{15}$ & 0.031  \\
$v_{23}$ & 0.031  \\
$v_{7}$ & 0.030  \\
$v_{10}$ & 0.030  \\
$v_{2}$ & 0.026  \\
$v_{21}$ & 0.025  \\
$v_{19}$ & 0.024  \\
$v_{4}$ & 0.023  \\
$v_{18}$ & 0.023  \\
$v_{13}$ & 0.022  \\
$v_{16}$ & 0.022  \\
$v_{27}$ & 0.020  \\
$v_{32}$ & 0.020  \\
$v_{8}$ & 0.020  \\
$v_{9}$ & 0.020  \\
$v_{17}$ & 0.019  \\
$v_{31}$ & 0.019  \\
$v_{24}$ & 0.019  \\
$v_{29}$ & 0.018  \\
$v_{30}$ & 0.017  \\
$v_{33}$ & 0.015  \\
$v_{14}$ & 0.015  \\
\hline
\end{tabular}
\caption{$\epsilon$ values obtained for $^{\mathrm{DIAG}}$EDC$_v^+$ simulations of 2-thiocytosine with one vibrational mode being deactivated and therefore using 32 normal modes. The $\epsilon$ values obtained when omitting a specific normal mode is listed next to the corresponding normal mode. The 33-D $^{\mathrm{DIAG}}$EDC$_v^+$ dynamics act as a reference for obtaining the $\epsilon$ values.}
\label{tab:2thio_singlemodes}
\end{table}

\begin{table}[b]
\begin{tabular}{ccc}
\hline
Model size& Vibrational modes & $\epsilon$ \\
\hline
10-D & $v_{28}$, $v_{20}$, $v_{26}$, $v_{1}$, $v_{22}$, $v_{25}$, $v_{5}$, $v_{12}$, $v_{6}$, $v_{3}$ & 0.219\\
9-D & $v_{28}$, $v_{20}$, $v_{26}$, $v_{1}$, $v_{22}$, $v_{25}$, $v_{5}$, $v_{12}$, $v_{6}$ & 0.222 \\
8-D & $v_{28}$, $v_{20}$, $v_{26}$, $v_{1}$, $v_{22}$, $v_{25}$, $v_{5}$, $v_{12}$ & 0.220 \\
7-D & $v_{28}$, $v_{20}$, $v_{26}$, $v_{1}$, $v_{22}$, $v_{25}$, $v_{5}$ & 0.225 \\
6-D & $v_{28}$, $v_{20}$, $v_{26}$, $v_{1}$, $v_{22}$, $v_{25}$  & 0.341\\
5-D & $v_{28}$, $v_{20}$, $v_{26}$, $v_{1}$, $v_{22}$ & 0.318\\
4-D & $v_{28}$, $v_{20}$, $v_{26}$, $v_{1}$ & 0.272 \\
3-D & $v_{28}$, $v_{20}$, $v_{26}$ & 0.640 \\
2-D & $v_{28}$, $v_{20}$ & 0.688 \\
1-D & $v_{28}$ & 0.772 \\
\hline
\end{tabular}
\caption{Employed lower-dimensional model systems for 2-thiocytosine ranging from 1 to 10-D. The included vibrational modes and the final $\epsilon$ values obtained for $^{\mathrm{DIAG}}$EDC$_v^+$ simulations with respect to the 33-D LVC model are listed as well.}
\label{tab:2thio_lowerd}
\end{table}

\clearpage

\subsection{Population dynamics with laser fields}
\begin{figure}[h]
\includegraphics{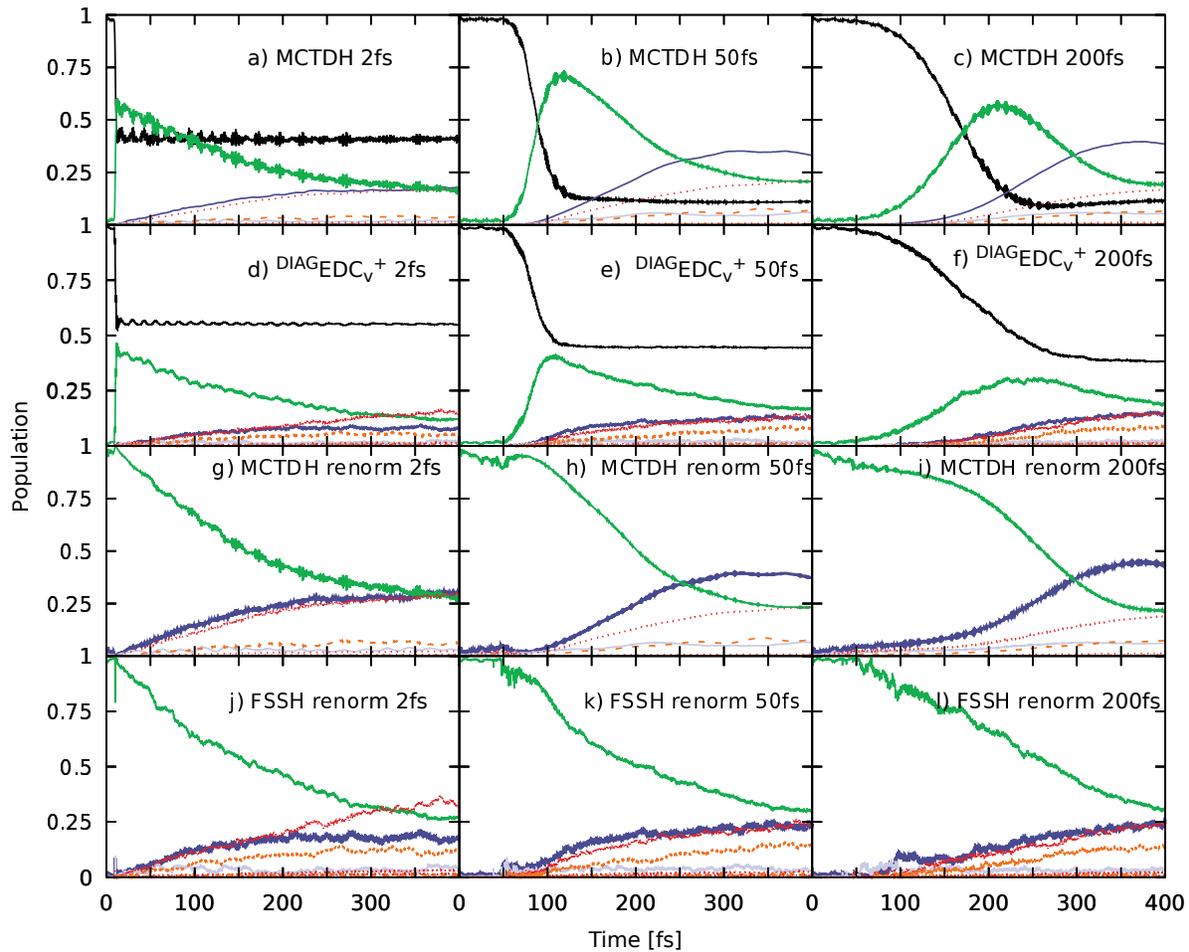}
\caption{\label{fig:2thio_laserpops} Evolution of the populations in all considered states in the 10-D model of the 2-thiocytosine molecule in presence of laser fields. a-c) MCTDH populations for dynamics using varying length ($t_P$) of the laser pulse. d-f) FSSH simulations using the $^{\mathrm{DIAG}}$EDC$_\mathrm{v}^+$ set of parameters for the same laser lengths already employed in a-c). g-i) Renormalized excited state populations for the MCTDH dynamics shown in a-c) where each excited state population is adapted according to $p^r_i=\frac{p_i}{1-p_1}$ where $p_1$ denotes the ground state population and $p_i$ the population in the $i$th excited state in a-c). j-l) Renormalized excited state populations for the FSSH dynamics shown in d-f).}
\end{figure}

The influence of employing longer laser pulses to excite 2-thiocytosine are shown in FIG~\ref{fig:2thio_laserpops} where the top two rows show the population evolution for MCTDH (first row) and the $^\mathrm{DIAG}$EDC$_\mathrm{v}^+$ set of parameters (second row).
The remaining columns show the influence of renormalization using the S$_0$ population in every time step (see Section~\ref{sec:so2_pop} for a more detailed explanation).

\clearpage

\subsection{$\epsilon$ values}

\begin{table}[b]
\scriptsize
\begin{tabular}{ccccc}
\hline
& $\epsilon$ & $\epsilon_{sing}$ & $\epsilon_{trip}$ & \\
\hline
$^\mathrm{MCH}$EDC$_g^{+}$      & 0.1637        & 0.1053        & 0.0584        & \\
$^\mathrm{MCH}$EDC$_g^g$        & 0.1654        & 0.1041        & 0.0613        & \\
$^\mathrm{DIAG}$NONE$_{v}^{v}$  & 0.1816        & 0.0950        & 0.0866        & \\
$^\mathrm{DIAG}$NONE$_{v}^{+}$  & 0.1974        & 0.1124        & 0.0850        & \\
$^\mathrm{MCH}$NONE$_{v}^{v}$   & 0.2113        & 0.1368        & 0.0745        & \\
$^\mathrm{DIAG}$EDC$_{v}^{+}$   & 0.2145        & 0.1584        & 0.0561        & \\
$^\mathrm{MCH}$NONE$_{v}^{+}$   & 0.2154        & 0.1318        & 0.0836        & \\
\hline
\end{tabular}
\caption{$\epsilon$ values obtained for all sets of considered FSSH dynamics in a 10-D model of 2-thiocytosine without any laser field. MCTDH used as reference. The methods are sorted according to lowest overall $\epsilon$.}
\label{tab:2thio_epsilons_tot}
\end{table}

\begin{table}
\scriptsize
\begin{tabular}{ccccc}
\hline
& $\epsilon$ & $\epsilon_{sing}$ & $\epsilon_{trip}$ & \\
\hline
$^\mathrm{DIAG}$NONE$_{v}^{v}$  & 0.1816        & 0.0950        & 0.0866        & \\
$^\mathrm{MCH}$EDC$_g^g$        & 0.1654        & 0.1041        & 0.0613        & \\
$^\mathrm{MCH}$EDC$_g^{+}$      & 0.1637        & 0.1053        & 0.0584        & \\
$^\mathrm{DIAG}$NONE$_{v}^{+}$  & 0.1974        & 0.1124        & 0.0850        & \\
$^\mathrm{DIAG}$NONE$_{none}^{+}$       & 0.2215        & 0.1302        & 0.0913        & \\
$^\mathrm{MCH}$NONE$_{v}^{+}$   & 0.2154        & 0.1318        & 0.0836        & \\
$^\mathrm{MCH}$NONE$_{v}^{v}$   & 0.2113        & 0.1368        & 0.0745        & \\
\hline
\end{tabular}
\caption{Identical information to TABLE~\ref{tab:2thio_epsilons_tot}. The methods are sorted according to lowest $\epsilon_{sing}$.}
\label{tab:2thio_epsilons_sing}
\end{table}

\begin{table}
\scriptsize
\begin{tabular}{ccccc}
\hline
& $\epsilon$ & $\epsilon_{sing}$ & $\epsilon_{trip}$ & \\
\hline
$^\mathrm{MCH}$NONE$_g^g$       & 0.3517        & 0.3073        & 0.0444        & \\
$^\mathrm{MCH}$NONE$_g^{+}$     & 0.3343        & 0.2862        & 0.0482        & \\
$^\mathrm{MCH}$AFSSH$_g^{+}$    & 0.3112        & 0.2599        & 0.0514        & \\
$^\mathrm{MCH}$EDC$_{v}^{+}$    & 0.2191        & 0.1668        & 0.0523        & \\
$^\mathrm{MCH}$EDC$_{v}^{v}$    & 0.2255        & 0.1705        & 0.0551        & \\
$^\mathrm{DIAG}$EDC$_{v}^{+}$   & 0.2145        & 0.1584        & 0.0561        & \\
$^\mathrm{DIAG}$EDC$_{v}^{v}$   & 0.2198        & 0.1617        & 0.0581        & \\
\hline
\end{tabular}
\caption{Identical information to TABLE~\ref{tab:2thio_epsilons_tot}. The methods are sorted according to lowest $\epsilon_{trip}$.}
\label{tab:2thio_epsilons_trip}
\end{table}

\begin{table}
\scriptsize
\begin{tabular}{cccccc}
\hline
& $\epsilon$ & $\epsilon_{S_0}$ & $\epsilon^r_{sing}$ & $\epsilon^r_{trip}$ & \\
\hline
$^\mathrm{MCH}$AFSSH$_g^{+}$    & 0.3763        & 0.2323        & 0.0664        & 0.0775        & \\
$^\mathrm{MCH}$AFSSH$_g^g$      & 0.3784        & 0.2334        & 0.0695        & 0.0755        & \\
$^\mathrm{MCH}$EDC$_g^g$        & 0.3859        & 0.2584        & 0.0841        & 0.0434        & \\
$^\mathrm{MCH}$EDC$_g^{+}$      & 0.3859        & 0.2590        & 0.0830        & 0.0439        & \\
$^\mathrm{DIAG}$EDC$_{none}^{+}$        & 0.4506        & 0.2396        & 0.1549        & 0.0561        & \\
$^\mathrm{MCH}$EDC$_{none}^{+}$ & 0.4920        & 0.2625        & 0.1708        & 0.0587        & \\
$^\mathrm{DIAG}$EDC$_{v}^{v}$   & 0.4944        & 0.2376        & 0.1704        & 0.0864        & \\
\hline
\end{tabular}
\caption{$\epsilon$ values obtained for all sets of considered FSSH dynamics in a 10-D of 2-thiocytosine including a $t_p=30$~fs laser field using MCTDH as reference. The methods are sorted according to lowest overall $\epsilon$}
\label{tab:2thio_epsilons_laser_tot}
\end{table}

\begin{table}
\scriptsize
\begin{tabular}{cccccc}
\hline
& $\epsilon$ & $\epsilon_{S_0}$ & $\epsilon^r_{sing}$ & $\epsilon^r_{trip}$ & \\
\hline
$^\mathrm{DIAG}$EDC$_h^{+}$     & 0.6932        & 0.2134        & 0.2644        & 0.2154        & \\
$^\mathrm{DIAG}$EDC$_h^h$       & 0.6785        & 0.2144        & 0.2561        & 0.2081        & \\
$^\mathrm{DIAG}$EDC$_g^{+}$     & 0.6844        & 0.2205        & 0.2584        & 0.2054        & \\
$^\mathrm{DIAG}$EDC$_g^g$       & 0.6696        & 0.2226        & 0.2486        & 0.1984        & \\
$^\mathrm{MCH}$AFSSH$_g^{+}$    & 0.3763        & 0.2323        & 0.0664        & 0.0775        & \\
$^\mathrm{MCH}$AFSSH$_g^g$      & 0.3784        & 0.2334        & 0.0695        & 0.0755        & \\
$^\mathrm{MCH}$AFSSH$_{v}^{v}$  & 0.5114        & 0.2363        & 0.1827        & 0.0924        & \\
\hline
\end{tabular}
\caption{Identical information to TABLE~\ref{tab:2thio_epsilons_laser_tot}. The methods are sorted according to lowest $\epsilon_{S_0}$.}
\label{tab:2thio_epsilons_laser_s0}
\end{table}

\begin{table}
\scriptsize
\begin{tabular}{cccccc}
\hline
& $\epsilon$ & $\epsilon_{S_0}$ & $\epsilon^r_{sing}$ & $\epsilon^r_{trip}$ & \\
\hline
$^\mathrm{MCH}$AFSSH$_g^{+}$    & 0.3763        & 0.2323        & 0.0664        & 0.0775        & \\
$^\mathrm{MCH}$AFSSH$_g^g$      & 0.3784        & 0.2334        & 0.0695        & 0.0755        & \\
$^\mathrm{MCH}$EDC$_g^{+}$      & 0.3859        & 0.2590        & 0.0830        & 0.0439        & \\
$^\mathrm{MCH}$EDC$_g^g$        & 0.3859        & 0.2584        & 0.0841        & 0.0434        & \\
$^\mathrm{MCH}$NONE$_g^g$       & 0.4969        & 0.3055        & 0.0985        & 0.0929        & \\
$^\mathrm{MCH}$NONE$_g^{+}$     & 0.5236        & 0.3090        & 0.1189        & 0.0957        & \\
$^\mathrm{DIAG}$NONE$_h^h$      & 0.5006        & 0.2681        & 0.1387        & 0.0938        & \\
\hline
\end{tabular}
\caption{Identical information to TABLE~\ref{tab:2thio_epsilons_laser_tot}. The methods are sorted according to lowest $\epsilon^r_{sing}$.}
\label{tab:2thio_epsilons_laser_sing}
\end{table}

\begin{table}
\scriptsize
\begin{tabular}{cccccc}
\hline
& $\epsilon$ & $\epsilon_{S_0}$ & $\epsilon^r_{sing}$ & $\epsilon^r_{trip}$ & \\
\hline
$^\mathrm{MCH}$EDC$_g^g$        & 0.3859        & 0.2584        & 0.0841        & 0.0434        & \\
$^\mathrm{MCH}$EDC$_g^{+}$      & 0.3859        & 0.2590        & 0.0830        & 0.0439        & \\
$^\mathrm{DIAG}$EDC$_{none}^{+}$        & 0.4506        & 0.2396        & 0.1549        & 0.0561        & \\
$^\mathrm{MCH}$EDC$_{none}^{+}$ & 0.4920        & 0.2625        & 0.1708        & 0.0587        & \\
$^\mathrm{MCH}$EDC$_{v}^{v}$    & 0.5180        & 0.2628        & 0.1902        & 0.0650        & \\
$^\mathrm{MCH}$EDC$_{v}^{+}$    & 0.5262        & 0.2646        & 0.1934        & 0.0681        & \\
$^\mathrm{DIAG}$AFSSH$_{none}^{+}$      & 0.5202        & 0.2805        & 0.1715        & 0.0682        & \\
\hline
\end{tabular}
\caption{Identical information to TABLE~\ref{tab:2thio_epsilons_laser_tot}. The methods are sorted according to lowest $\epsilon^r_{trip}$.}
\label{tab:2thio_epsilons_laser_trip}
\end{table}
\clearpage

\bibliography{dynlib.bib}